\documentclass[aps,prd, floatfix,notitlepage,nofootinbib,twocolumn,superscriptaddress]{revtex4-1}
\usepackage{amsmath}
\usepackage{amssymb}
\usepackage{graphicx}
\usepackage[tight]{subfigure}
\usepackage{enumitem}
\usepackage{soul}
\usepackage{ulem}
\usepackage{pifont}
\usepackage{xspace}

\makeatletter
\renewcommand{\p@subsection}{}
\renewcommand{\p@subsubsection}{}
\makeatother

\usepackage[english]{babel}
\makeatletter
\def\bbl@set@language#1{%
  \edef\languagename{%
    \ifnum\escapechar=\expandafter`\string#1\@empty
    \else\string#1\@empty\fi}%
  \@ifundefined{babel@language@alias@\languagename}{}{%
    \edef\languagename{\@nameuse{babel@language@alias@\languagename}}%
  }%
  \select@language{\languagename}%
  \expandafter\ifx\csname date\languagename\endcsname\relax\else
    \if@filesw
      \protected@write\@auxout{}{\string\select@language{\languagename}}%
      \bbl@for\bbl@tempa\BabelContentsFiles{%
        \addtocontents{\bbl@tempa}{\xstring\select@language{\languagename}}}%
      \bbl@usehooks{write}{}%
    \fi
  \fi}
\newcommand{\DeclareLanguageAlias}[2]{%
  \global\@namedef{babel@language@alias@#1}{#2}%
}
\makeatother
\DeclareLanguageAlias{en}{english}

\usepackage{bm}

\usepackage[pdfusetitle,colorlinks=true,citecolor=blue,linkcolor=magenta]{hyperref}

\graphicspath{{./}{./images/}}

\begin{document}

\title{Emergent conformal symmetry in non-unitary random dynamics of free fermions}

\author{Xiao Chen}
\affiliation{Department of Physics, Boston College, Chestnut Hill, MA 02467, USA}
\affiliation{Department of Physics and Center for Theory of Quantum Matter, University of Colorado, Boulder, CO 80309, USA}
\author{Yaodong Li}
\affiliation{Department of Physics, University of California, Santa Barbara, CA 93106, USA}
\author{Matthew P. A. Fisher}
\affiliation{Department of Physics, University of California, Santa Barbara, CA 93106, USA}
\author{Andrew Lucas}   
\affiliation{Department of Physics and Center for Theory of Quantum Matter, University of Colorado, Boulder, CO 80309, USA}

\date{\today}

\begin{abstract}
We present random quantum circuit models for non-unitary quantum dynamics of free fermions in one spatial dimension.  Numerical simulations reveal that the dynamics tends towards steady states with logarithmic violations of the entanglement area law and power law correlation functions.  Moreover, starting with a short-range entangled many-body state, the dynamical evolution of entanglement and correlations quantitatively agrees with the predictions of two-dimensional conformal field theory with a space-like time direction.
We argue that this behavior is generic in non-unitary free
quantum dynamics with time-dependent randomness, and show that the emergent conformal dynamics of two-point functions arises out of a simple ``nonlinear master equation".
\end{abstract}

\maketitle


\section{Introduction}
Recent years have seen a surge of interest in many-body quantum dynamics generated by random unitary circuits \cite{lashkari_towards_2013, nahum_2017, nahum_operator_2018,rakovszky_diffusive_2017,von_keyserlingk_operator_2018,khemani_operator_2017,zhou_operator_2018,xu_locality_2018,Gopalakrishnan2018,Bentsen2019,chen_power_law_2019,Pai2019,Iaconis2019,kn, Zhou_Nahum_2019,Rakovszky_diffusive_2019,Chang_2019, Sarang_Austen_2019,chen2019quantum}.
These models are simplified cartoons for the unitary quantum dynamics of many-body systems, and allow for numerical or even analytic descriptions of the physics of thermalization and dissipation,  operator growth and many-body chaos, entanglement spreading, and diffusion. 
As the dynamics of these systems is unitary and highly chaotic, the endpoint of the dynamical evolution is typically a thermalized state with volume law entanglement, and remains as featureless as possible given the symmetries of the model.


The story qualitatively changes if the quantum dynamics is not unitary, where the emergent steady states need not be thermal and featureless, and can exhibit interesting and unexpected structures.
An example of this dynamics is a random unitary circuit subject to random projective measurements \cite{Cao_Tilloy_2019,Skinner_2019,Li_2018,Chan_2019}.
In this system, there is a phase transition: the entanglement entropy remains volume law at slow measurement rate, and enters an area law phase at fast measurement rate \cite{Skinner_2019, Li_2018, Chan_2019, gullans2019dynamical, gullans2019scalable, zabalo2019critical, choi2019quantum, Tang_Zhu_2020, Li_2019, Szyniszewski_2019, zhang2020nonuniversal, goto2020measurementinduced, jian2019measurementinduced, bao2019theory,fan2020selforganized}.
At the critical point, analytical and numerical results provide strong evidence for emergent conformal symmetry \cite{Skinner_2019,Li_2019,Li_2020}. Note that to observe this transition, we need to follow the {\it quantum trajectory} of the many-body wave function rather than the evolution of the density matrix described by the Kraus map \cite{kraus1971general} or its Markovian version, the Lindblad equation \cite{lindblad1976}.

Motivated by these studies, in this paper we introduce a model of random non-unitary dynamics for free fermions.
Our model consists of discrete time evolution, with alternating application of unitary gates (nearest-neighbor hopping gates), non-unitary gates (evolving with on-site potential in imaginary time), and wave function renormalization.
This model is different than a free fermion model subjected to projective measurement \cite{Chan_2019,Cao_Tilloy_2019}, in which any non-zero measurement rate drives the system to a trivial quantum Zeno phase with area law entanglement.
In our model, based on extensive numerical simulations, we argue that so long as the model has time-dependent randomness, there is emergent  spacetime conformal symmetry \cite{francesco2012conformal} in the disorder-averaged (variance of) two-point functions, and in von Neumann entanglement entropy as well as the mutual information, regardless of the strength of the non-unitary gates.
More precisely, we conjecture that after $T \gg 1$ steps of the dynamics, the quantum state $|\psi(T)\rangle$ evolves to 
\begin{align}
|\psi(T)\rangle \approx \frac{\mathrm{e}^{-  T H_{\rm CFT}} |\psi_0\rangle }{\lVert \mathrm{e}^{-T H_{\rm CFT}}  |\psi_0\rangle \rVert}.
\label{eq:conj}
\end{align}
where $H_{\rm CFT}$ represents a CFT Hamiltonian in $1+1$d, whose precise form we do not know.
Reminiscent of self-organized critical systems \cite{Bak_1987,bak2013nature},  there is no finely-tuned parameter at criticality; these critical phenomena are remarkably robust to various perturbations and modifications of the model.
Due to the simplicity of free fermion dynamics, we hope that this model will be a useful starting point for a broader understanding emergent scale and even conformal invariance in non-unitary dynamics. 

The rest of the paper is organized as follows.
In Sec. \ref{sec:model}, we introduce the model and what we are going to compute.
In Sec. \ref{sec:steady} , we numerically study the properties of the steady states which arise when the number of time steps $T\gg L$, where $L$ is the system size.
In Sec. \ref{sec:dynamics}, we explore the time evolution starting from a short range entangled state, and understand the crossover from $T<L$ to $T>L$.
In Sec. \ref{sec:master_eq}, we provide an interpretation of this emergent critical dynamics, and argue for its robustness, by deriving a ``nonlinear master equation" for the two-point correlation functions in a continuous time model with Brownian non-unitary dynamics.
In Sec. \ref{sec:discussion}, we summarize our results and discuss several interesting directions for future work.  


\section{The model and the method}
\label{sec:model}
\begin{figure}[t]
\centering
\includegraphics[width=\columnwidth]{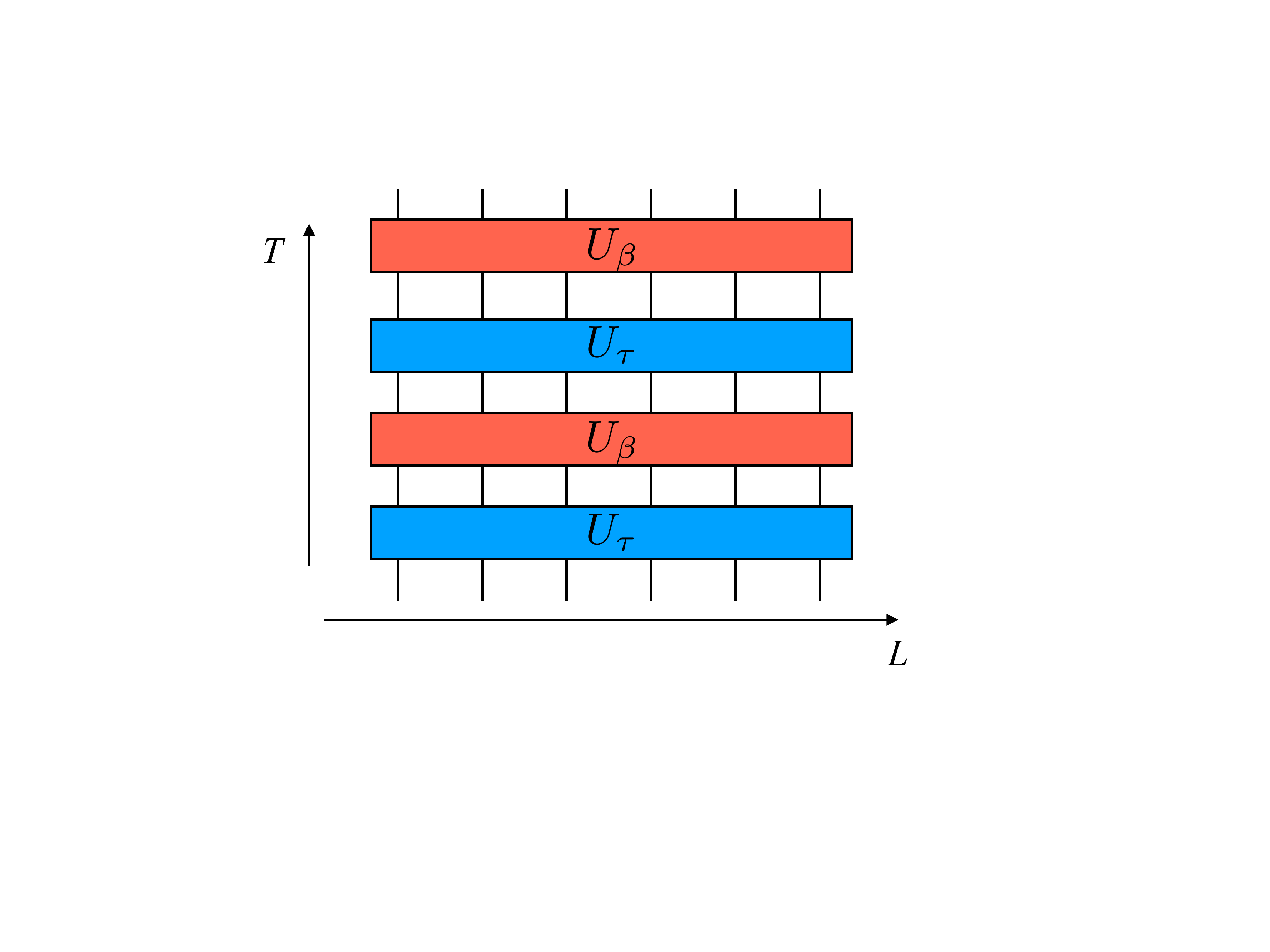}
\caption{The schematics for the non-unitary random dynamics of free fermions.}
\label{fig:schematics}
\end{figure}

\begin{figure*}[t]
\centering
\subfigure[]{
  \label{fig:Corr_collapse_beta_0}
  \includegraphics[width=.8\columnwidth]{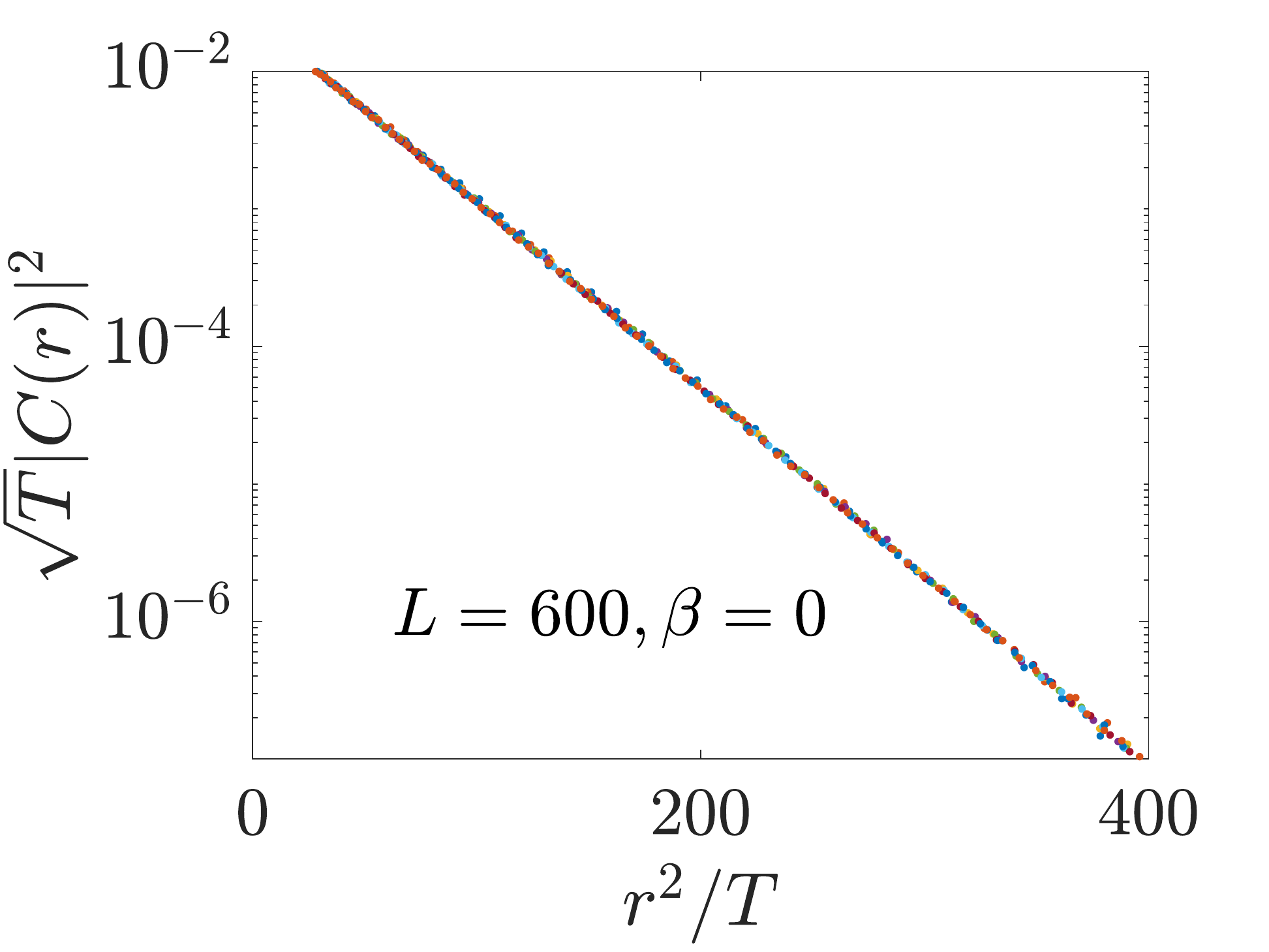}
}
\subfigure[]{
  \label{fig:S_T_beta_0}
  \includegraphics[width=.8\columnwidth]{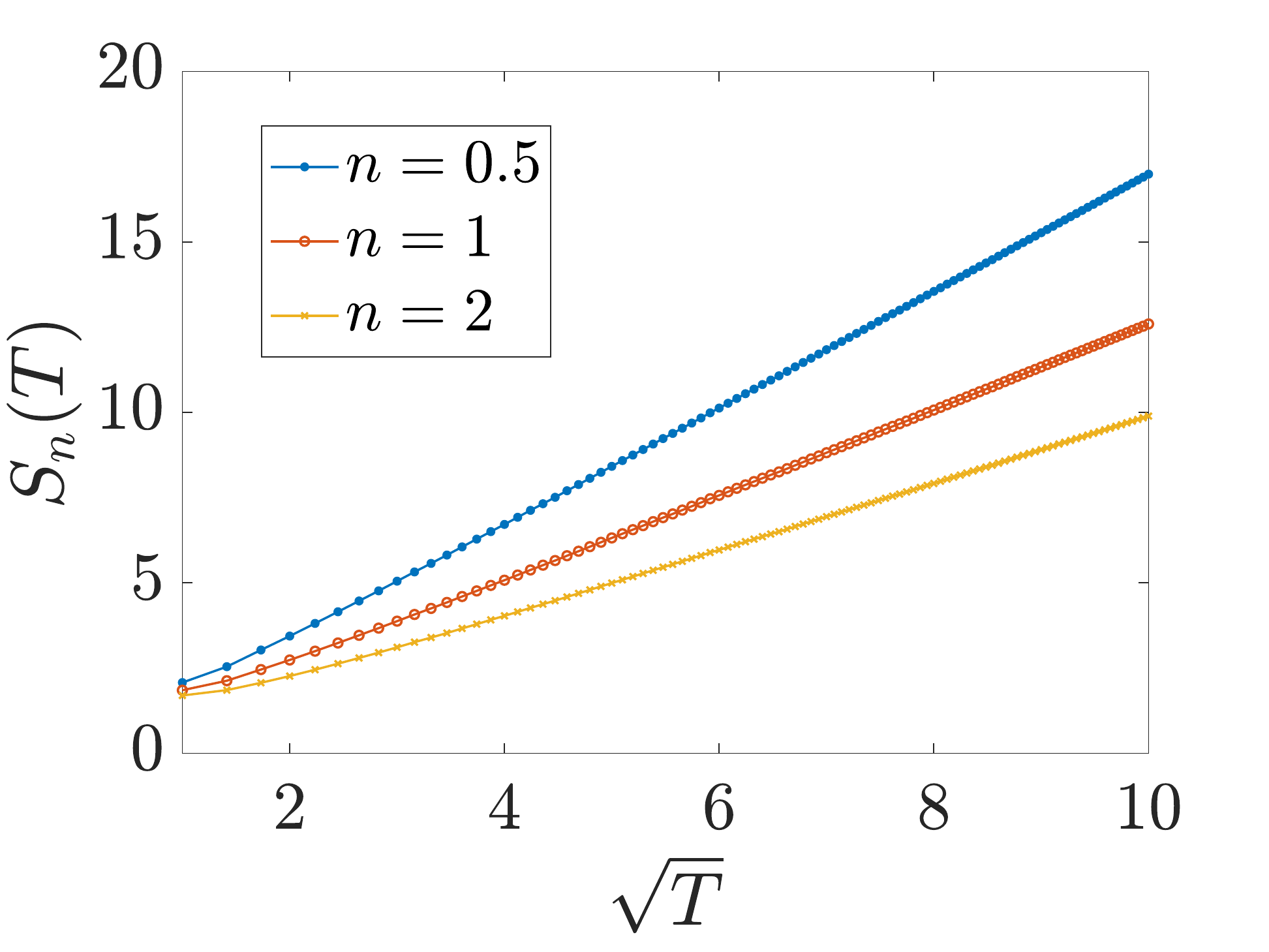}
}
\caption{The quantum dynamics at $\beta=0$ with open boundary condition where $L$ is the system size. (a) is the data collapse of squared correlation function between $T\in[20,100]$. Here $C(r)\equiv C_{L/2-x,L/2+x+1}$ with $r=2x+1$. All the data collapse into a single curve. (b) is the growth of R\'enyi entanglement entropy for half of the system vs $\sqrt{T}$.}
\label{fig:beta_0}	
\end{figure*}

In this section, we consider non-unitary random dynamics with the time evolution operator (shown in Fig.~\ref{fig:schematics})
\begin{align}
U=\prod_{t=1}^{T} U_\beta(t)U_\tau(t),
\label{eq:non_unitary}
\end{align}
which consists of both unitary and imaginary time evolutions.
Here $U_\tau(t)=\exp(-2\mathrm{i}\tau H_1(t))$ denotes unitary evolution for time $\tau$ for a one dimensional fermionic chain with random nearest neighbor hopping.
In the simulations we discuss below, the Hamiltonian $H_1(t)$ is a tight binding model, defined as
\begin{align}
H_1(t)= \sum_{x}\kappa_{x,t}c^{\dag}_xc_{x+1}+\mathrm{H.c.}
\label{eq:h_1}
\end{align}
The second part of each period $U_\beta(t)=\exp(-2\beta H_2(t))$ denotes imaginary time evolution for an ``imaginary time unit'' $\beta$, where the Hamiltonian $H_2(t)$ is a simple random onsite potential:
\begin{align}
H_2(t)=\sum_{x}\lambda_{x,t}c^{\dag}_{x}c_{x}.
\end{align}
Both $H_1$ and $H_2$ are random in both space and time.
The parameters $\kappa_{x,t}$ and $\lambda_{x,t}$ are independent random variables with a distribution $P_\kappa(\kappa_{x,t})$ and $P_\lambda(\lambda_{x,t})$. Specifically, we take a simple two-component distribution,
\begin{align}
&P_\kappa(\kappa_{x,t})=p_1\delta(\kappa_{x,t}-1)+(1-p_1)\delta(\kappa_{x,t}+1)\\
&P_\lambda(\lambda_{x,t})=p_2\delta(\lambda_{x,t}-1)+(1-p_2)\delta(\lambda_{x,t})
\end{align}
with $p_1, p_2\in [0,1] $.

We are interested in the wave function dynamics,
\begin{align}
|\psi(T)\rangle=\frac{U(T)}{\sqrt{Z}}|\psi_0\rangle.
\end{align} 
where $Z=\langle \psi_0| U^\dag(T)U(T)|\psi_0\rangle$.
The initial pure state $|\psi_0\rangle$ is chosen to be a short-ranged entangled state:  \begin{equation}
    |\psi_0\rangle = |\cdots 01010101\cdots\rangle. \label{eq:psi0neel}
\end{equation}
Under time evolution with $H_1$ and $H_2$ chosen as above, $|\psi(T)\rangle$ remains a fermionic Gaussian state \cite{bravyi2004lagrangian}; therefore, the entire state is fully encoded in the two point correlation matrix $C(T)$, with
\begin{equation}
    \label{eq:C_matrix_def}
    C_{xy}(T) \equiv \langle \psi(T)|  c^\dag_xc_y|\psi(T)\rangle.
\end{equation}
Numerical algorithms to compute the evolution of $C(T)$ are explained in Appendix \ref{app:C_algorithm}.
We also observe from  (\ref{eq:C_matrix_def}) that $C$ is a projection operator satisfying 
\begin{align}
\mbox{Tr}C=\mbox{Tr}C^2=N,
\label{eq:projection}
\end{align}
where $N$ is the number of particles and is conserved under non-unitary time evolution.
(\ref{eq:projection}) is an important identity which we will use later.
Given $C(T)$, we can further compute the entanglement entropy for a subsystem.  This is because $|\psi(T)\rangle$ is a Gaussian state and satisfies Wick's theorem \cite{Peschel_2003}.
For von Neumann entanglement entropy in particular, we have
\begin{align}
S_{\rm vN}=-\mathrm{Tr}\left[C_A\log C_A + (1-C_A)\log(1-C_A) \right], \label{eq:SvNn1}
\end{align}
where $C_A$ is the correlation matrix defined in the subsystem A.
We can further compute the generalized R\'enyi entropy:
\begin{align}
S_{n}=\frac{1}{1-n}\mathrm{Tr}\log\left[C_A^n + (1-C_A)^n \right]. \label{eq:SRenyi}
\end{align}
where $n$ is the R\'enyi index.
In the limit $n\rightarrow 1$, (\ref{eq:SRenyi}) reduces to (\ref{eq:SvNn1}).

Before we analyze the non-unitary dynamics, we briefly discuss the simplest case with $\beta=0$. This corresponds to the unitary time evolution. For the random dynamics described by (\ref{eq:h_1}), we expect to observe diffusive dynamics \cite{Roosz2016}.
We numerically confirm this result and present it in Fig. \ref{fig:beta_0}.
In Fig. \ref{fig:Corr_collapse_beta_0}, we show that $C_{x,x+r}$ spreads out diffusively, i.e., 
\begin{align}
    \overline{|C_{x,x+r}|^2}\sim \frac{e^{-r^2/T}}{\sqrt{T}},
\end{align}
and the averaged R\'enyi entropies also exhibit diffusive scaling, $\overline{S_n}\sim \sqrt{T}$,  regardless of the R\'enyi index $n$ (see Fig. \ref{fig:S_T_beta_0}). $\overline{S_n}$ will saturate to volume law after sufficient time evolution.
Throughout the paper, $\overline{S_n}$ and $\overline{|C_{x,x+r}|^2}$ are numerically obtained through ensemble averaging over different circuit realization (as specified by $\{\kappa_{x,t}\}$ and $\{\lambda_{x,t}\}$). Therefore we may drop the overline frequently in the rest of the paper. 

\section{Steady state}
\label{sec:steady}
\begin{figure*}[hbt]
\centering
\subfigure[]{
  \label{fig:Corr_p1_05_p2_05}
  \includegraphics[width=.8\columnwidth]{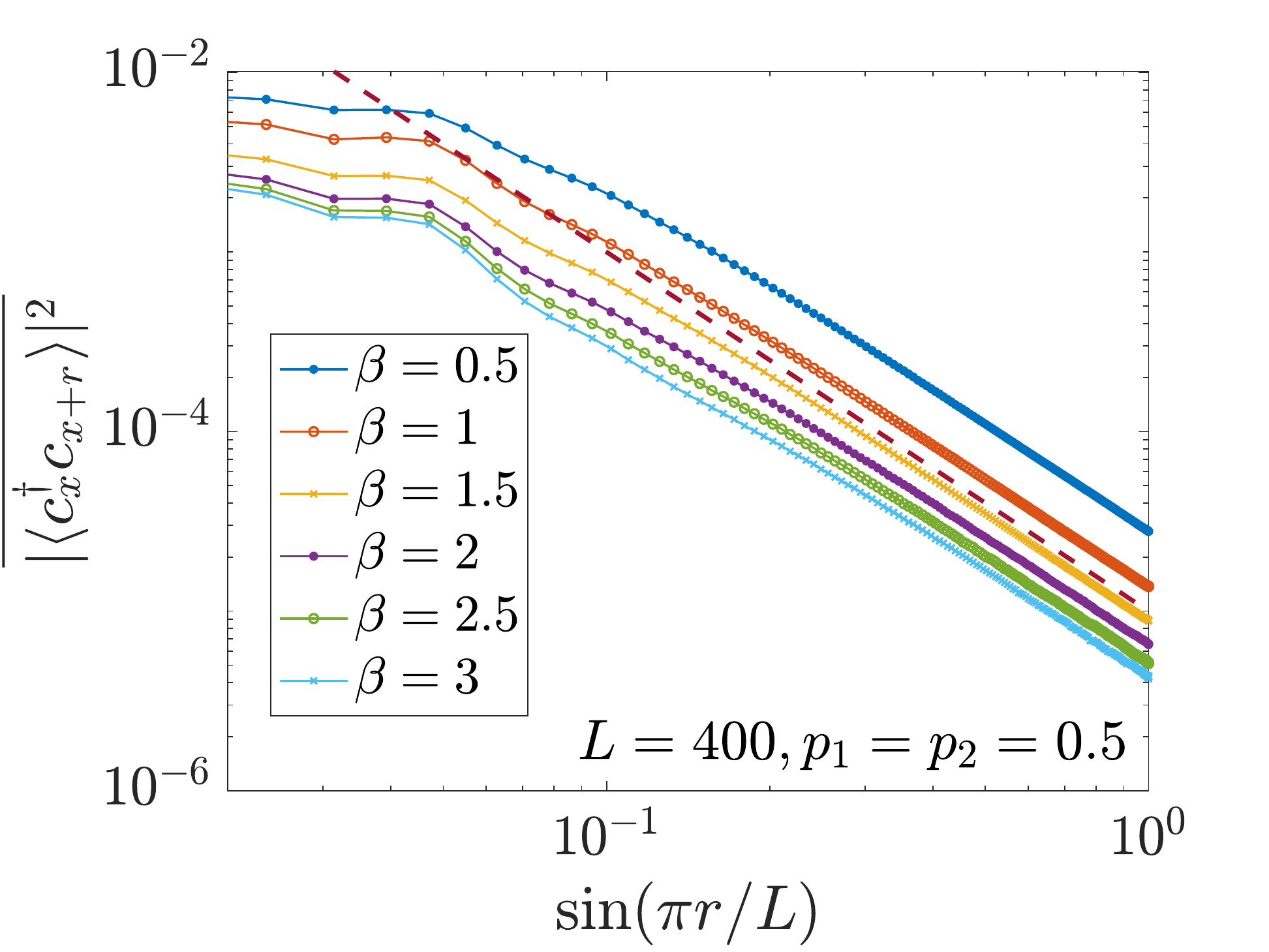}
}
\subfigure[]{
  \label{fig:S1_LA_p1_05}
  \includegraphics[width=.8\columnwidth]{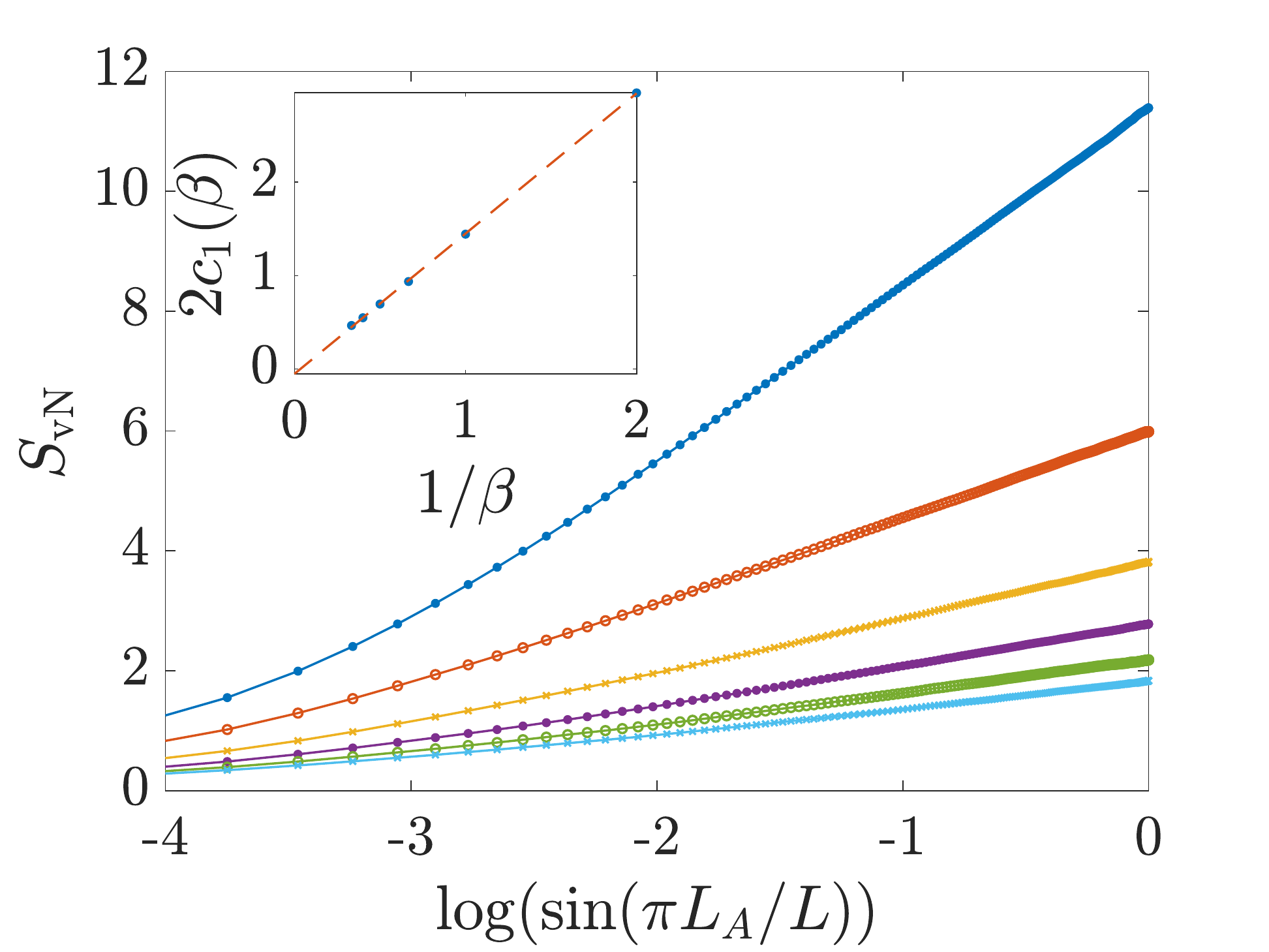}
}
\subfigure[]{
  \label{fig:MI_eta_p1_05_p2_05}
 \includegraphics[width=.8\columnwidth]{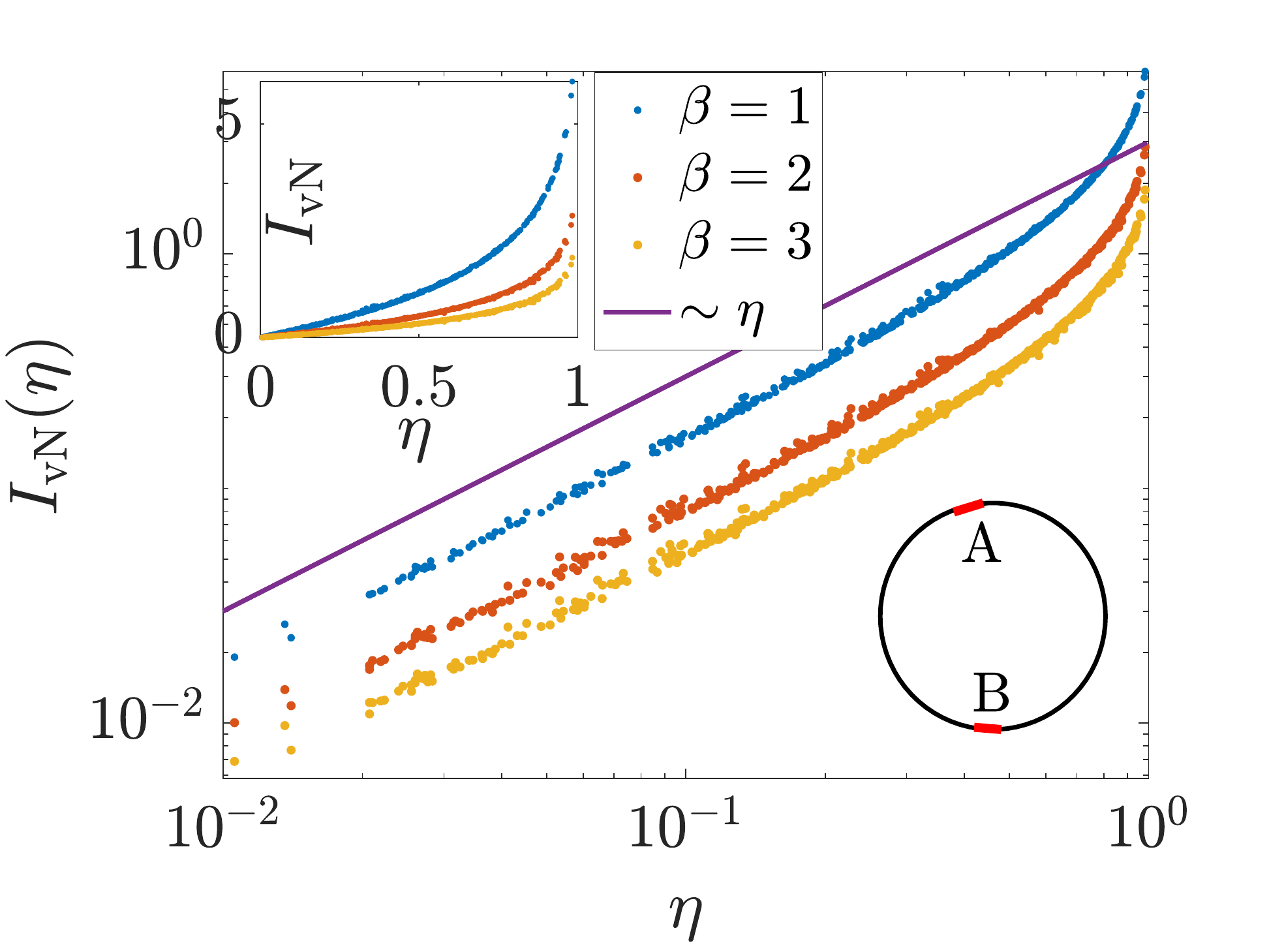}
}
\subfigure[]{
  \label{fig:S1_LA_filling}
 \includegraphics[width=.8\columnwidth]{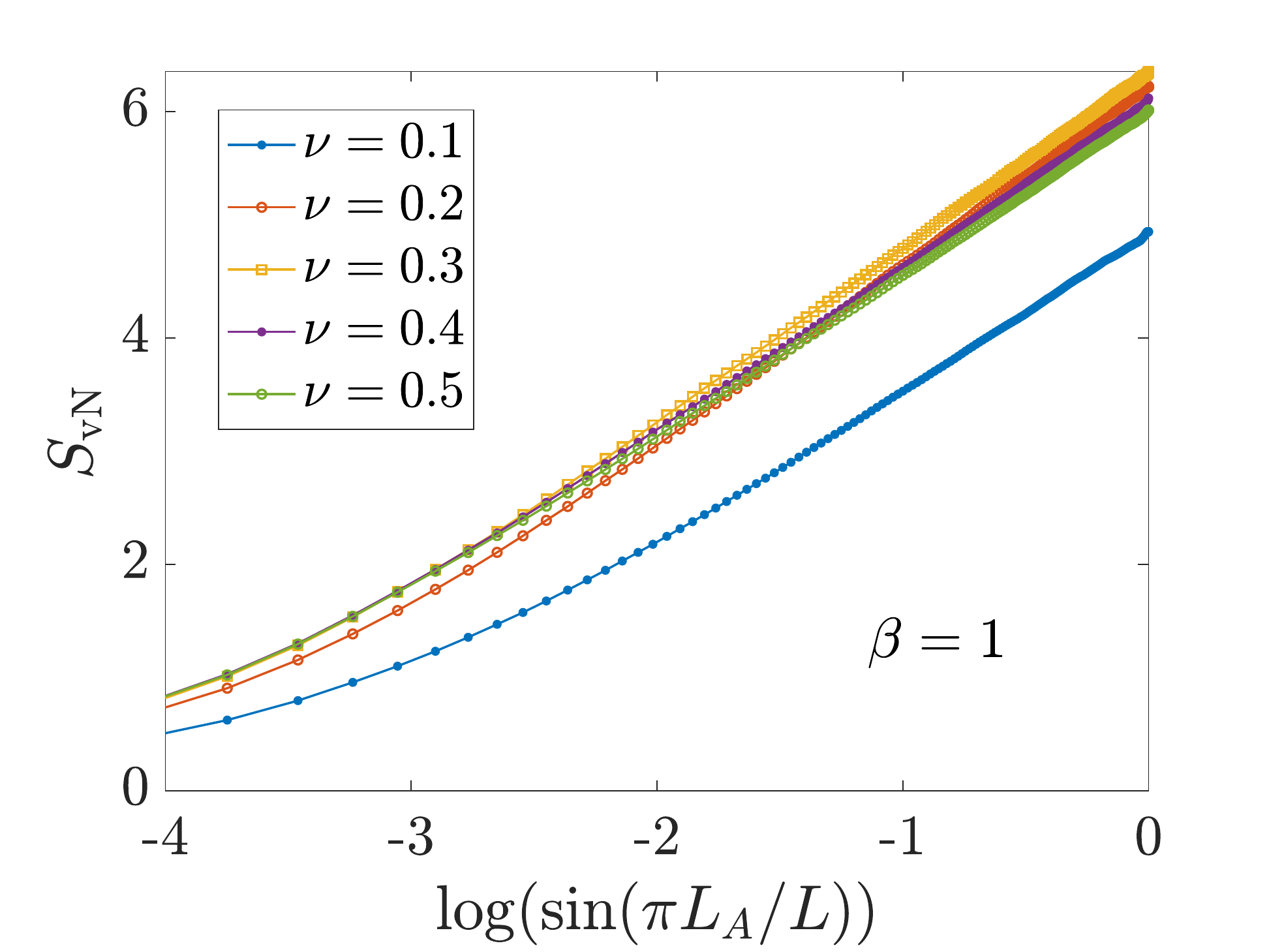}
}
\caption{The numerical results of the steady state for various $\beta$ at $p_1=p_2=0.5$ with $L=400$ and periodic boundary condition. For the first three plots, the filling factor is fixed at $1/2$. (a) Squared correlation function on the log-log scale. The slope of the curves is 2 and is the same   as the dashed line which scales as $1/(\sin(\pi r/L))^2$. (b) von Neumann entanglement entropy $S_{\rm vN}$ vs $\log(\sin(\pi L_A/L))$ on the linear scale. The $\beta$ of the curve is the same as that in (a). The coefficient $2c_1(\beta)$ vs $1/\beta$ is shown in the inset. (c) The mutual information $I_1$ as a function of the cross ratio $\eta$ on the log-log scale. The same data is plotted in the inset on the linear scale. The intervals $A=[x_1,x_2]$ and $B=[x_3,x_4]$. The locations of $x_i$ are chosen randomly on the circle with the constraint $|x_i-x_j|>3$. (d) $S_{\rm vN}$ vs $\log(\sin(\pi L_A/L))$ at different filling factor $\nu=N/L$.}
\label{fig:p1_05_p2_05}	
\end{figure*}

First we characterize the steady state in the limit $T\to\infty$ for a one dimensional system with $L$ sites and periodic boundary conditions.
We fix $\tau=1$, and vary both $\beta$ (the imaginary time unit) and $\nu=N/L$ (the filling fraction).
Since this is a random system, the averaged two point correlation function 
$\overline{C_{x,x+r}}=\overline{\langle c_x^\dag c_{x+r}\rangle}=0$.
On the other hand, as shown in Fig.~\ref{fig:Corr_p1_05_p2_05}, the averaged squared correlation function $\overline{|\langle c_x^\dag c_{x+r}\rangle|^2}$  (i.e. the second moment) is nonzero \cite{Li_2019}.
Numerics shows that for $r\gg 1$, \begin{equation}
    \overline{|\langle c_x^\dag c_{x+r}\rangle|^2} \sim \frac{1}{r^2}. \label{eq:steadystater2}
\end{equation}
This power law scaling behavior indicates that this wave function is critical at finite $\beta$.
Furthermore, we find that the averaged von Neumann entanglement entropy of a set $A$ of $L_A$ adjacent sites scales as $\log(\sin(\pi L_A/L))$ with periodic boundary condition (see Fig.~\ref{fig:S1_LA_p1_05}). We further calculate the R\'enyi entanglement entropy and find results that are consistent with  
\begin{align}
S_n= c_1\left(1+\frac{1}{n}\right)\log \left[\frac{L}{\pi}\sin(\frac{\pi L_A}{L})\right].
\end{align}
This dependence on R\'enyi index is the same as the results of the ground state for a 1+1 dimensional conformal field theory (CFT) computed from the Cardy-Calabrese formalism \cite{Calabrese_2004,Calabrese_2009} (see Sec.~\ref{sec:dynamics} and Appendix.~\ref{app:CC_formalism} for detailed discussions).
The coefficient $c_1$ depends on $\beta$; numerically we find that when $p_1=p_2=0.5$, \begin{equation}
    c_1(\beta) \propto \frac{1}{\beta};
\end{equation}see the inset of Fig.~\ref{fig:S1_LA_p1_05}.

In addition, we compute the mutual information $I_n(A,B)=S_n(A)+S_n(B)-S_n(A\cup B)$ between two disjoint intervals $A=[x_1,x_2]$ and $B=[x_3,x_4]$, whose system sizes and locations can be varied.
We present the results in Fig.~\ref{fig:MI_eta_p1_05_p2_05} and we find that all the data points collapse to a single curve as a function of the cross ratio $\eta$, which is defined as
\begin{align}
\eta\equiv \frac{x_{12}x_{34}}{x_{13}x_{24}},\ {\rm with}\  x_{ij}=\sin\left(  \frac{\pi}{L}|x_i-x_j| \right).
\end{align}
Furthermore, $I(A,B)\propto \eta$ when $\eta \to 0$.
This limit can be taken by fixing $L_A = x_{12}$ and $L_B = x_{34}$, while taking the distance between $A$ and $B$ ($|x_{13}|$) to be large;
in this case, $\eta\sim |x_{13}|^{-2}$.
Therefore this result indicates that the mutual information between two small intervals scales as $1/r^2$ when their separation $r$ is large. This power law scaling is the same as that for the squared correlation function, consistent with the information-theoretic bound on $\overline{|\langle c_x^\dag c_{x+r}\rangle|^2}$~\cite{Wolf2008}.

The above critical scaling behavior also works at other filling factor (See Fig.~\ref{fig:S1_LA_filling}) and other values of $p_1$ and $p_2$.  These additional numerical results can be found in Appendix \ref{app:more_num}, and confirm that the emergent conformal symmetry is not finely tuned. We also consider the dimerized Hamiltonian for the unitary part in Eq.\eqref{eq:h_1} with even and odd bonds having different bond strength and we still observe the same critical behavior.

\section{Dynamics}
\label{sec:dynamics}
\begin{figure*}[hbt]
\centering
\subfigure[]{
  \label{fig:Corr_early}
  \includegraphics[width=.8\columnwidth]{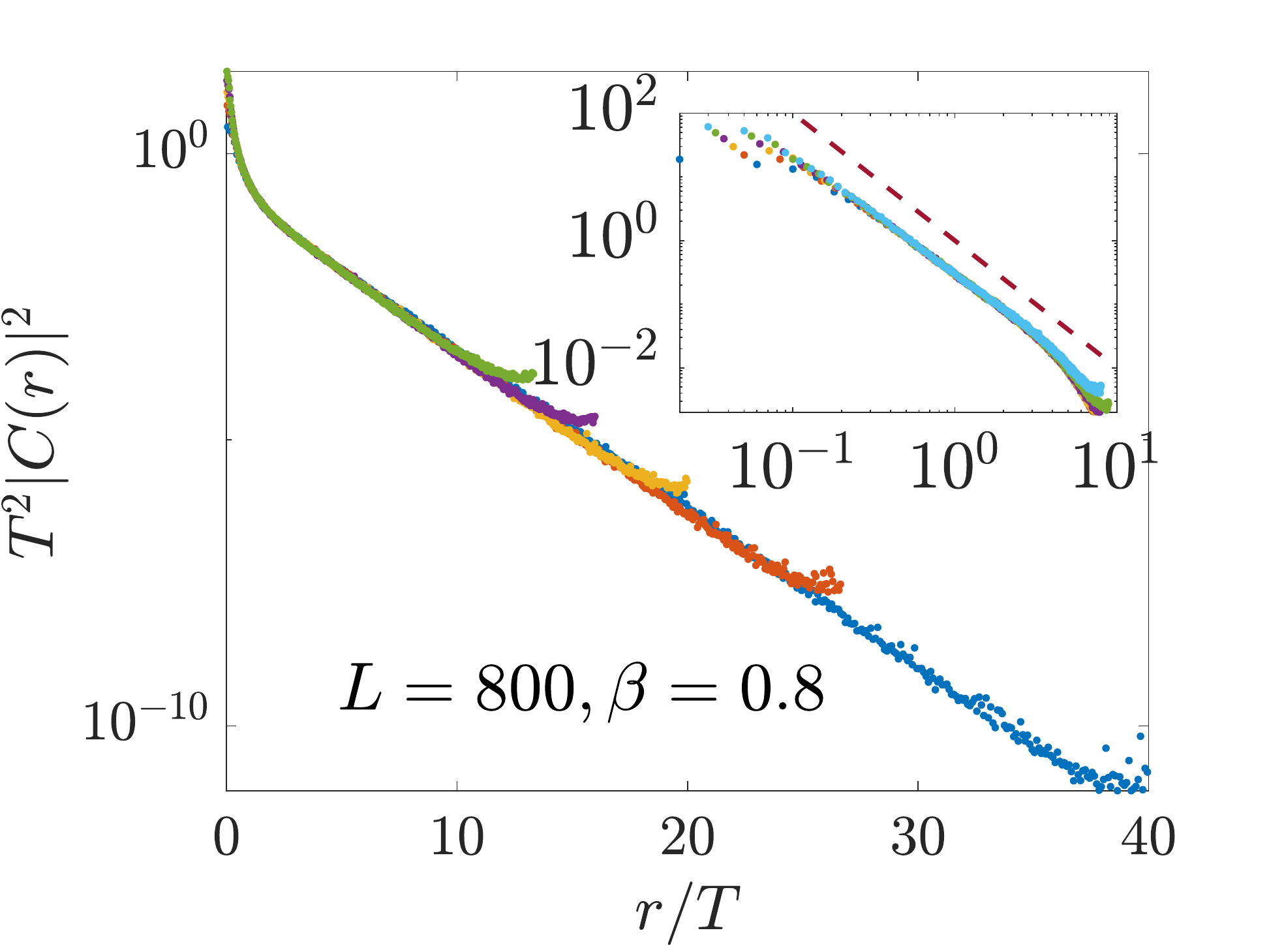}
}
\subfigure[]{
  \label{fig:S_half_T}
  \includegraphics[width=.8\columnwidth]{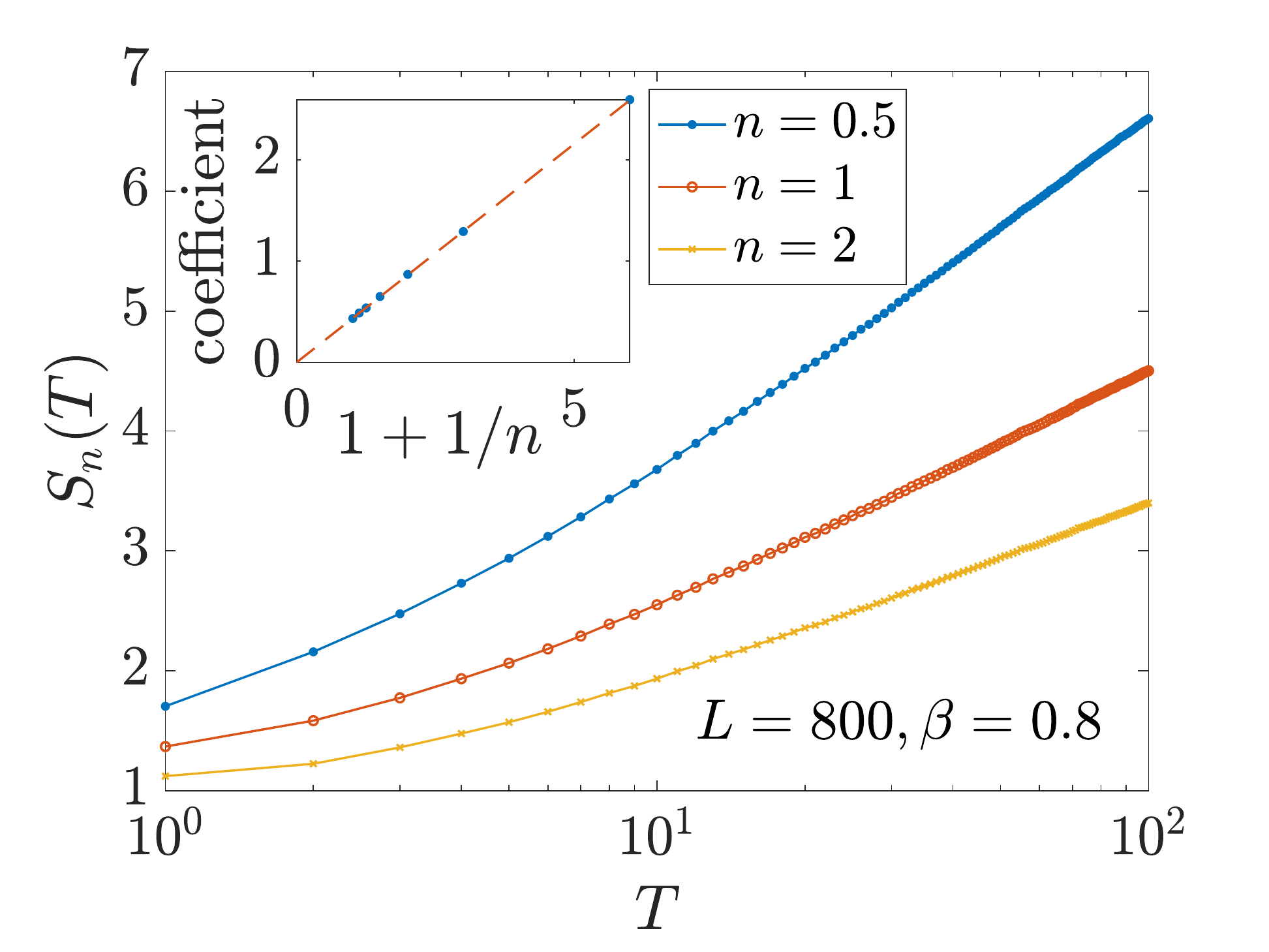}
}
\subfigure[]{
  \label{fig:S_collapse_beta}
 \includegraphics[width=.8\columnwidth]{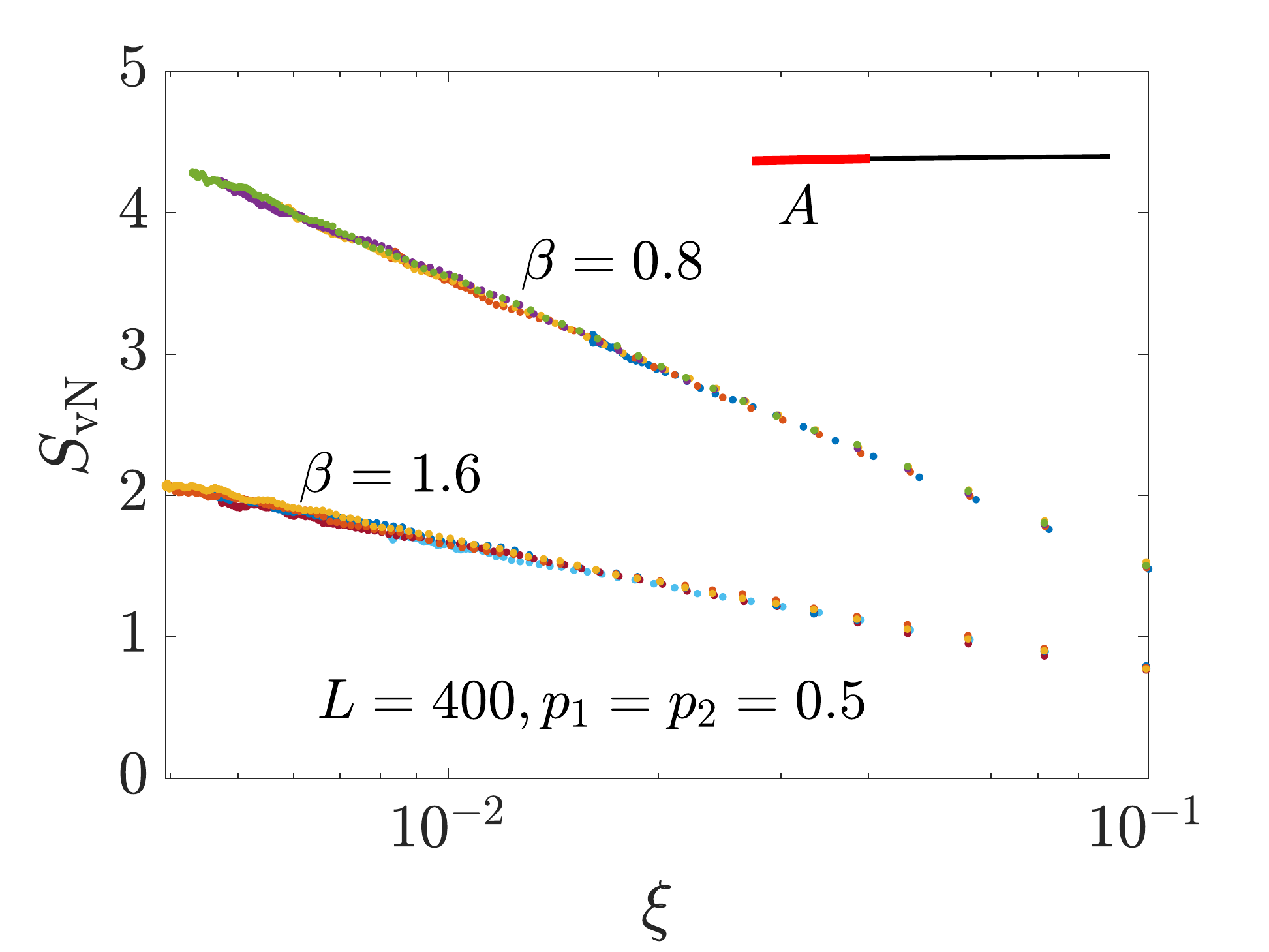}
}
\subfigure[]{
  \label{fig:MI_dyn_collapse}
 \includegraphics[width=.8\columnwidth]{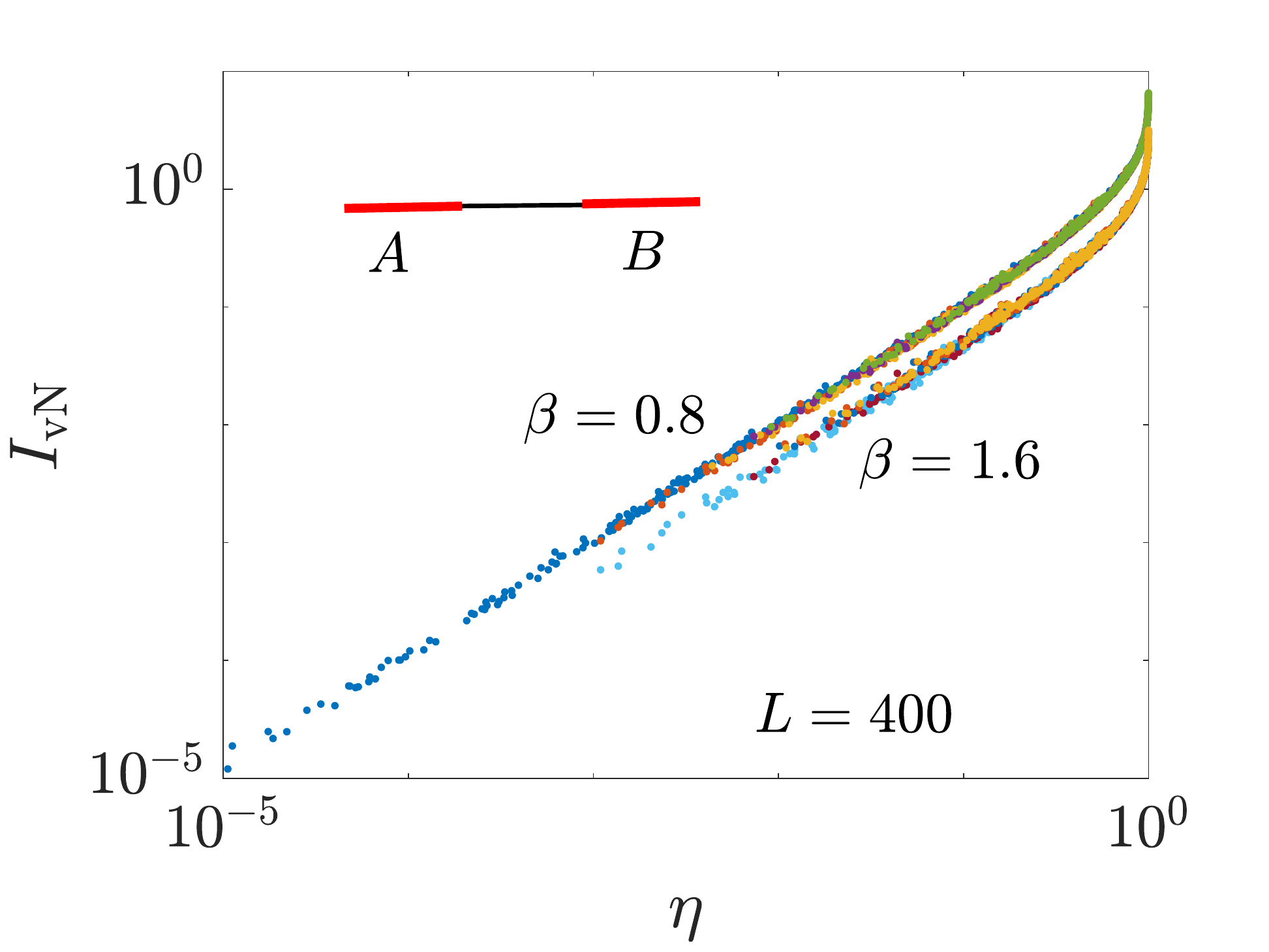}
}
\caption{(a) The data collapse of the squared correlation function at time $T\in[20, 60]$ on the semi-log scale. Here $C(r)\equiv C_{L/2-x,L/2+x+1}$ with $r=2x+1$. In the inset, we plot the same quantity on the log-log scale with $T\in [50, 100]$. The dashed line scales as $(T/2x+1)^2$. (b) The dynamics of $S_n$ for half of system on the semi-log scale. In the inset, we show the coefficient in front of $\log T$ vs $1+1/n$, where we take $n\in[0.2, 100]$. (c) The data collapse of entanglement dynamics for subsystem $A$ on the semi-log scale. $\xi$ depends on $L$, $L_A$ and $T$. All the data points of the same $\beta$ at different time collapse into a single straight curve. For both curves, the time $T\in[20, 100]$. For $\beta=0.8$, we choose the parameter $a=2.4$ (the detail can be found in Eq.\eqref{eq:aspect_ratio} in Appendix \ref{app:CC_formalism}), while for $\beta=1.6$, we take $a=4.8$.  (d) The data collapse of the mutual information dynamics between $A$ and $B$ for the time $T\in[20, 100]$. The other parameters are the same as in (c). For all the four plots, we consider open boundary condition with $p_1=p_2=0.5$.}
\label{fig:p1_05_p2_05}	
\end{figure*}

To better understand the physics of this model, we now explore the evolution of $|\psi(T)\rangle$ both when $T\ll L$ and $T\gg L$.  When $T\ll L$, as shown in Fig.~\ref{fig:Corr_early}, we find that when $T\ll r$, there exists a constant $a$ such that
\begin{align}
\overline{|\langle c_x^\dag c_{x+r}\rangle|^2}\sim \frac{\mathrm{e}^{-ar/T}}{T^2}.
\label{eq:early_time}
\end{align}
Due to the imaginary time evolution $U_\beta(t)$, the dynamics is no longer strictly local.
At early times, the correlation between two points decays exponentially in space with a correlation length proportional to time $T$.
This result holds in the thermodynamic limit, for  an arbitrarily large value of $r/T$.

We emphasize that this is not simply a mild breakdown of locality in the spirit of the Lieb-Robinson theorem \cite{Lieb1972} (which only guarantees an \textit{approximate} light cone for continuous unitary dynamics).
The discrete time unitary dynamics has an exact light cone \cite{nahum_operator_2018} which is destroyed specifically by the non-unitary dynamics.
Further discussion can be found in Appendix \ref{app:nolinearlightcone}.

In \eqref{eq:early_time}, the $r/T$ scaling suggests that the emergent criticality found previously has dynamical exponent $z=1$: namely, time and space scale together, as expected for a CFT.    As time evolves, we may write \begin{equation}
    \label{eq:scaling_function}
    \overline{|\langle c_x^\dag c_{x+r}\rangle|^2} = F\left(\frac{r}{T}\right) \frac{1}{T^2}.
\end{equation}
For large $x$, $F(x) \sim \exp(-ax)$; for small $x$, $F(x)\sim x^{-2}$.  Therefore at late times, we recover (\ref{eq:steadystater2}). Based on these numerical results, we conjecture that this non-unitary dynamics has emergent two-dimensional conformal symmetry: namely, the state is obtained through CFT Hamilontian under purely imaginary time evolution, as in (\ref{eq:conj}).

Following this assumption, it is easy to understand \eqref{eq:early_time}.
From the form of (\ref{eq:conj}), $T$ plays the role of an effective \textit{inverse temperature};  in a CFT this also serves as the correlation length.
This exponential decay behavior becomes algebraic when $T\gg r$, since on these length scales the physics is captured by the algebraic scaling of the ground state.

Next, we explore the growth of entanglement.  Again we consider a domain $A$ consisting of $L_A$ adjacent sites, and compute its entanglement entropy with the remaining sites as a function of time. When $T\ll L_A$, we find behavior consistent with 
\begin{align}
S_n=\frac{c_1}{2}\left(1+\frac{1}{n}\right) \log T.
\label{eq:EE_time}
\end{align} 
The $1/2$ prefactor is caused by open boundary conditions \cite{Calabrese_2004,Bastianello_2019}.\footnote{Here we switch from periodic boundary condition in the previous section to open boundary condition in this section.}
This result is consistent with the prediction from CFT \cite{Calabrese_2004,Calabrese_2009}, in which the calculation of the entanglement entropy is mapped to evaluate the correlation function for twist fields on a semi-infinite strip; see Appendix \ref{app:CC_formalism}.

Our model has a rectangular geometry ($T$ time steps and $L$ lattice sites), and in the numerical simulation, $L$ and $T$ are finite.
In a CFT, 
{we expect that}
\begin{align}
\label{eq:S_rec}
S_n=-\frac{c_1}{2}\left(1+\frac{1}{n}\right)\log\xi,
\end{align}
where $\xi$ depends on $L$, $T / L$ and $L_A / L$; the explicit formula is given in Appendix \ref{app:CC_formalism}. 
 
In Fig.~\ref{fig:S_collapse_beta}, we plot $S_{\rm vN}$ at different times as a function of $\xi$ and we find that all the data points collapse into a single straight curve, which provides strong numerics evidence that our conjecture in \eqref{eq:conj} is correct.
In addition, we further compute the mutual information dynamics for two intervals sitting next to the corner and we find that it is a function of cross ratio (see Fig.~\ref{fig:MI_dyn_collapse}), which is defined as
\begin{align}
\eta\equiv \frac{ |w_1-\overline{w}_1||w_2-\overline{w}_2|}{|w_1-\overline{w}_2||w_2-\overline{w}_1|},
\end{align}
 with $w_1$ and $w_2$ as functions of the parameters $L_A,L_B, L, T$ (The details of $w_1$ and $w_2$ can be found in Appendix \ref{app:CC_formalism}). 
Notice that when $\eta$ is close to $0$, we have $I_{\rm vN}\sim \eta$. The power law exponent is the same as that for the steady state with periodic boundary condition. In terms of Cardy-Calabrese formalism, which is used to compute the entanglement properties in CFT, the mutual information is related with the four point correlation function of the twist field \cite{Calabrese_2004,Calabrese_2009,Calabrese_MI_2009,Calabrese_MI_2011}. Therefore in our circuit model, the power law exponent 1 in the small $\eta$ expansion can be interpreted as the lowest scaling dimension of the allowed operators in the operator product expansion (OPE) of the twist field \cite{Calabrese_MI_2011}.

\section{Continuous time model}
\label{sec:master_eq}
In this section, we provide an alternative understanding of this critical behavior. Notice that this non-unitary random free fermion dynamics is Markovian and has the conservation law $\mbox{Tr}C^2=N$. This motivates us to write down a master equation to describe the spreading of the correlation function and its final steady state. To derive this master equation, we consider a continuous time model of non-unitary dynamics with Brownian noise, which we believe characterizes the same physics in the above discrete circuits.  As before, we consider free fermions on a one dimensional lattice of $L$ sites.   The instantaneous Hamiltonian is given by
\begin{align}
    \mathrm{d}H(t) &= \sum_j \left(c^\dagger_{j+1}c_j \mathrm{d}W_j(t) + c^\dagger_j c_{j+1} \mathrm{d}\overline{W}_j(t) \right. \notag \\
    &\;\;\;\;\; \left.- \mathrm{i}c^\dagger_j c_j \mathrm{d}W^\prime_j(t)\right), \label{eq:browniandH}
\end{align}
with $\mathrm{d}W_j$, $\mathrm{d}\overline{W}_j$ and $\mathrm{d}W_j^\prime$ representing three different Brownian motions.  

As we have discussed in Eq.\eqref{eq:projection}, the matrix of two-point functions $C$ completely characterizes a state of free fermions.
It is a projection matrix, which satisfies $\mbox{Tr}C^2=N$.
For an initial product state, only the diagonal elements are non-zero.
As time evolves, the off-diagonal elements also becomes non-zero, while maintaining the same trace constraint.
This motivates us to define a distribution function $f_n$ which captures the spreading of ``weight" in the $C$ matrix:
\begin{align}
    f_n\equiv 
    \begin{cases}
    \sum_a \frac{|C_{a,a}|^2}{N}, &\mbox{when}\ n=0\\
    \sum_{a}\frac{|C_{a,a+n}|^2+|C_{a,a-n}|^2}{N}, & \mbox{when}\ n>0
    \end{cases},
    \label{eq:fndef}
\end{align}


\subsection{Nonlinear master equation}
Our goal is to derive an (approximate) equation governing the dynamics of $f_n(t)$.
This technical computation is given in Appendix \ref{app:mastereqn}.
The result is 
 \begin{subequations} \label{eq:master}\begin{align}
\partial_t f_1 &=  \mu + \theta (f_2-2f_1) - 2f_1 \sum_{m=1}^\infty f_m  \notag \\
&+ \sum_{m=1}^\infty f_m f_{m+1}, \\
\partial_t f_n &= \theta(f_{n+1}+f_{n-1}-2f_n) - 2f_n \sum_{m=1}^\infty f_m \notag \\
&+ \sum_{m=1}^\infty f_m f_{m+n} + \frac{1}{2}\sum_{m=1}^{n-1}f_m f_{n-m}, \;\;\; (n>1)
\end{align}\end{subequations}
Here $\mu$ and $\theta$ are positive constants.  $\theta$ is large when the amplitude of the unitary nearest-neighbor hopping is much larger than the amplitude of the non-unitary on-site term; this limit is analogous to the $\beta \rightarrow 0$ limit in the discrete time circuit.  We have set the ``strength" of the non-unitary terms to 1 in our effective master equation, and have rescaled time.  The degree of freedom $f_0$ is unique in that its average value must be non-zero, since (\ref{eq:projection}) holds at all times.  For this reason we ignore it in our approximate master equation;  justification for this is provided in Appendix \ref{app:mastereqn}.  
This derivation is not mathematically exact, but as we will show below, this set of equations exhibits the same emergent conformal symmetry that we saw before, and we believe that these master equations capture the key physics of the Brownian model (\ref{eq:browniandH}), and more generally of our non-unitary random free fermion dynamics.

Remarkably, time-independent solutions to (\ref{eq:master}) are known analytically \cite{KRAPIVSKY1993157}, and take the form $f_n \propto n^{-2}$ (at large $n$). However, at finite time $t$ and for our initial condition $f_n(t)=0$, an exact solution is not known.  We propose a self-consistent solution to the system (\ref{eq:master}) of the form \begin{equation}
    f_n(t) \approx t^{-\alpha} F(\varphi), \;\;\; (n\gg 1) \label{eq:ansatz}
\end{equation} 
where \begin{equation}
    \varphi = \frac{n}{t}.
\end{equation}
Notice the resemblance between this ansatz and the scaling form in \eqref{eq:scaling_function}.
This solution will be valid on times $t\gg 1$, and on this time scale $f_m$ are approximately time-independent for $m\sim 1$.   Upon plugging this ansatz in to (\ref{eq:master}), we obtain the following heuristic equation \begin{align}
    &-\frac{\alpha F(\varphi)}{t^{\varphi+1}} - \frac{\varphi F^\prime(\varphi)}{t^{\alpha+1}} \approx \frac{\varphi F^{\prime\prime}(\varphi)}{t^{\alpha+2}} \notag \\
    &+ \int\limits_{1/t}^{\varphi/2} \mathrm{d}\zeta \frac{F(\zeta)}{t^{2\alpha-1}}\left(F(\varphi+\zeta)+F(\varphi-\zeta)-2F(\varphi)\right). \label{eq:heuristicmaster}
\end{align}
The integration limits are not exact, but do capture the dominant terms in the equation.
The first observation is that at large $t$, the $\theta$ term is always subleading; hence we may ignore this contribution.  Physically, this means that the unitary dynamics is actually \textit{irrelevant} for maintaining the shape of the distribution at late times!   Next, observe that when $\alpha<2$, the convolution term dominates at large $t$;  by dominant balance there must be something equally large to balance this term, and so $\alpha\ge 2$.   On the other hand, if $\alpha > 2$, the convolution term is irrelevant;  were this the case, then we could exactly solve the diffusion equation $\partial_t f \approx \zeta \partial_n^2 f$ and our scaling ansatz (\ref{eq:ansatz}) would be wrong.  We conclude that $\alpha=2$ if the non-unitary dynamics plays any non-trivial role.

Next we analyze $F(\varphi)$ when $\varphi \ll 1$, where (\ref{eq:heuristicmaster}) reads \begin{equation}
    -2F - \varphi F^\prime=  \int\limits_{1/t}^{\varphi/2} \mathrm{d}\zeta F(\zeta)\left(F(\varphi+\zeta)+F(\varphi-\zeta)-2F(\varphi)\right). \label{eq:heuristicmaster2}
\end{equation}
Suppose the right hand side could be ignored;  if it could, then \begin{equation}
    F(\varphi) \sim \varphi^{-2} \;\;\; (\varphi \ll 1).  \label{eq:Fxim2}
\end{equation}
If the right hand side is not vanishing, then $F(\varphi)$ must decrease faster than in (\ref{eq:Fxim2}).  Suppose that $F(\zeta) \sim \zeta^{-\gamma-2}$ with $\gamma \ge 0$, as $\zeta \rightarrow 0$.  Then if $\gamma\ge 1$, the convolution is dominated by $\varphi \sim 1/t$, and the equation is not time-dependent;  so we may take $\gamma<1$, in which case we crudely estimate that \begin{equation}
    \int\limits_{1/t}^{\varphi/2} \mathrm{d}\zeta F(\zeta)\left(F(\varphi+\zeta)+F(\varphi-\zeta)-2F(\varphi)\right) \sim \varphi^{1-\gamma} F^{\prime\prime}(\varphi). \label{eq:convolutionestimate}
\end{equation}
As $\varphi \rightarrow 0$, this term is always subleading; we conclude that (\ref{eq:Fxim2}) holds.

When $\varphi\gg 1$, it is difficult to explicitly solve (\ref{eq:heuristicmaster}) because the convolution term is not even approximately local.  However, our argument in (\ref{eq:convolutionestimate}) still gives insight:  if $F(\varphi)$ was a power law at large $\varphi$, then we would be able to estimate the convolution term as quasi-local, and we would obtain a small correction to the equation of motion.  We would then find $F(\varphi)\approx \varphi^{-2}$ at all $\varphi$, and a time-independent $f_n(t)$!  That is one schematic solution to the equations of motion, but not the one we are after -- it is already in steady-state!   The other possibility is that \begin{equation}
    F(\varphi) \approx \mathrm{e}^{-\varphi}.
\end{equation}
In this case, the convolution in (\ref{eq:heuristicmaster2}) balances the derivative contribution, and the $-2F$ term in (\ref{eq:heuristicmaster2}) is subleading.   This solution does exhibit non-trivial dynamics, and describes the dynamical evolution of the distribution to its steady-state.

To summarize, we have given a heuristic argument that the non-unitary free fermion Brownian dynamics is well-captured by the nonlinear master equation (Eq.\eqref{eq:master}), which in turn exhibits the scaling solution \begin{equation}
    f_n(t) \sim \left\lbrace\begin{array}{ll} n^{-2} &\ n\ll t \\ t^{-2}\exp(-n/t) &\ n\gg t \end{array}\right.. \label{eq:heuristicconclusion}
\end{equation}
This precisely agrees with the predictions of CFT and of (\ref{eq:conj}), as discussed above.

Let us quickly note that slightly similar equations have appeared in the literature before under the name of ``aggregation dynamics" \cite{krapivsky}.  In the simplest of these equations, the infinite sums  in (\ref{eq:master}) are absent, and this qualitatively changes the dynamics.  The precise form of the nonlinear convolution terms in (\ref{eq:master}) is crucial to see emergent criticality.

\subsection{Numerical simulations}
\begin{figure*}[hbt]
\centering
\subfigure[]{
  \label{fig:f_early_collapse}
  \includegraphics[width=.65\columnwidth]{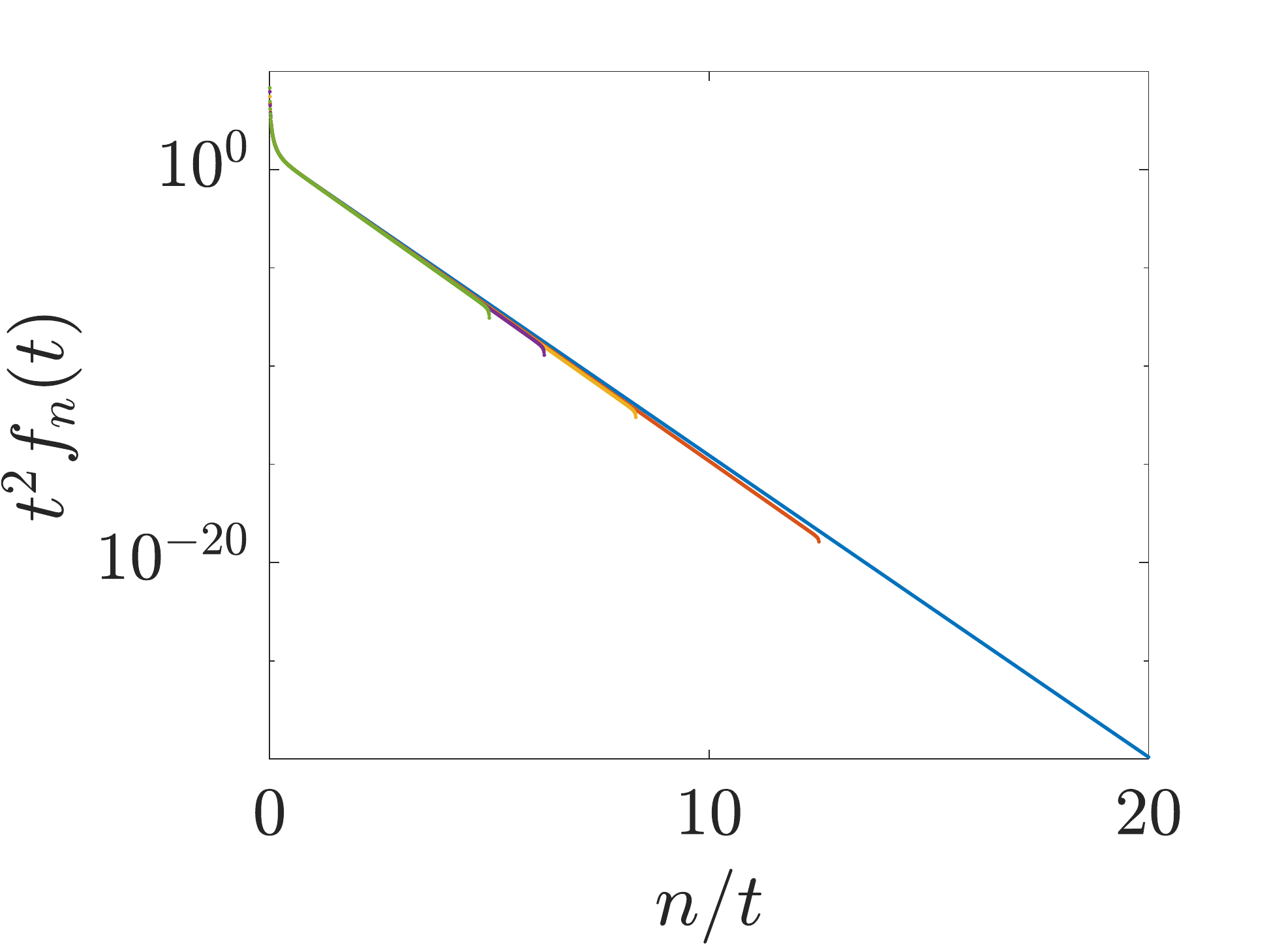}
}
\subfigure[]{
  \label{fig:f_late_collapse}
 \includegraphics[width=.65\columnwidth]{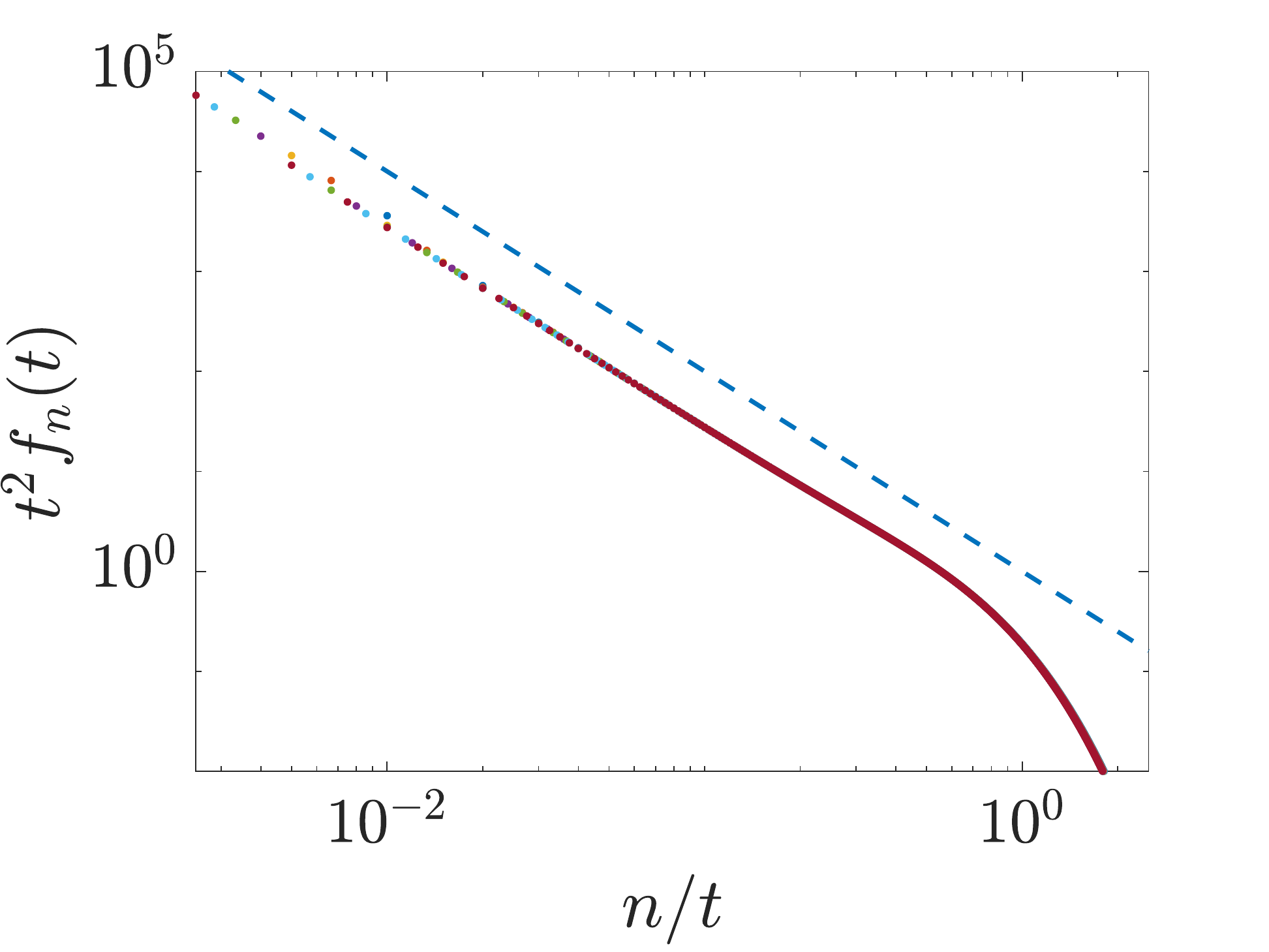}
}
\subfigure[]{
  \label{fig:sum_f}
 \includegraphics[width=.65\columnwidth]{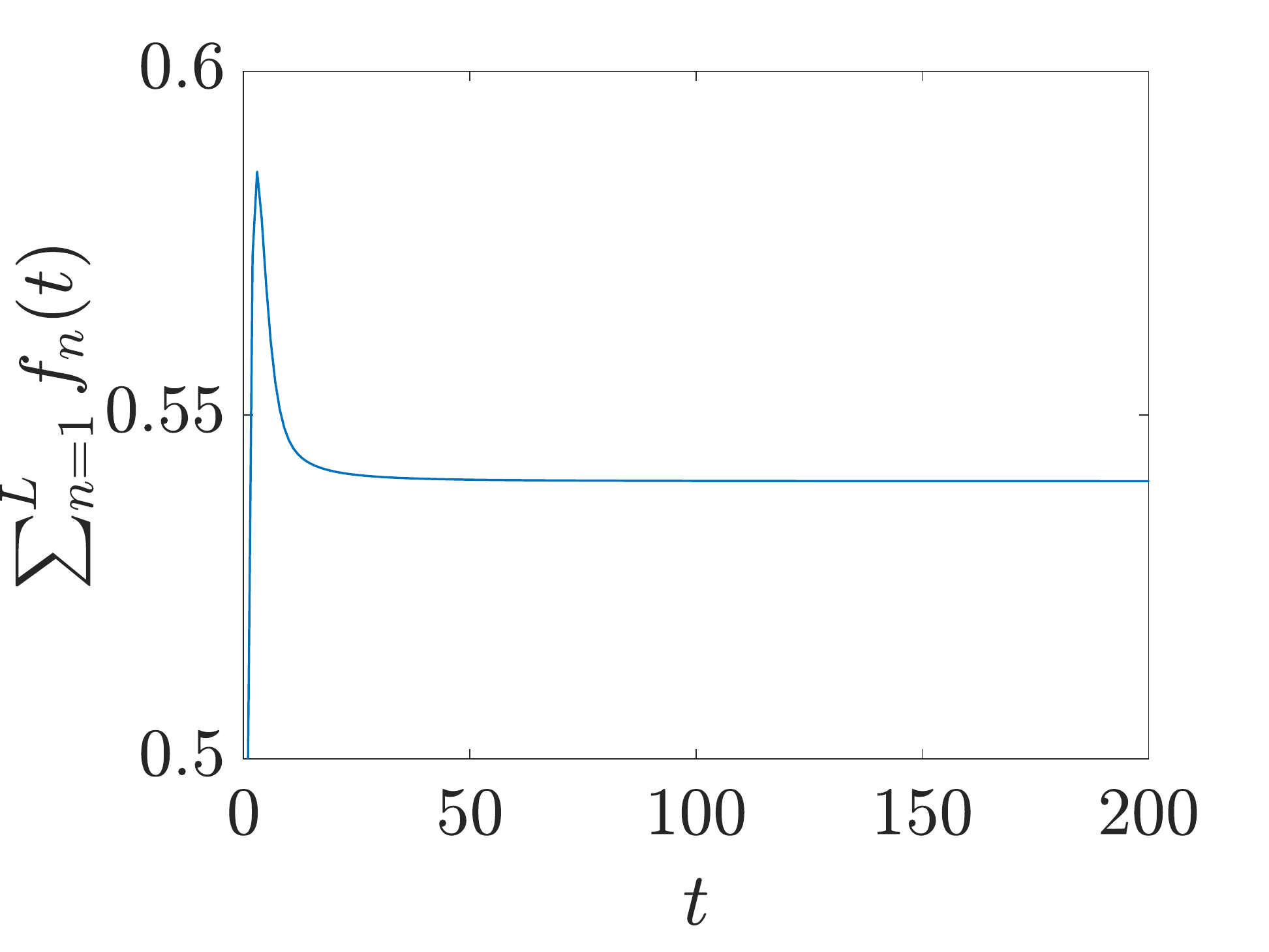}
}
\caption{The numerical results for the master equation in Eq.\eqref{eq:master} with $L=1000$ and $\theta=0$. (a) Data collapse of $t^2f_n(t)$ vs $t/n$ on the semi-log scale with $T\in[40,200]$. (b) Data collapse of $t^2f_n(t)$ vs $t/n$ on the log-log scale with $T\in[100,400]$. The dashed line scales as $(t/n)^2$. (c) The sum of $f_{n\geq 1}(t)$ vs time, which quickly approaches a constant, as demanded by the exact microscopic equations of motion.}
\label{fig:master_eq}	
\end{figure*}

We now confirm (\ref{eq:heuristicconclusion}) in explicit simulations of (\ref{eq:master}).  Since the diffusion term is not important for the late time dynamics, we set $\theta=0$.
Our results are shown in Fig.~\ref{fig:master_eq}, where we take $\mu=1$ and the initial condition $f_{n\geq 1}=0$. The data collapse in Fig.~\ref{fig:f_early_collapse} and \ref{fig:f_late_collapse}  indicate that $t^2f_n(t)$ is a function of $n/t$, consistent with the ansatz proposed in \eqref{eq:ansatz}. Furthermore, $t^2f_n(t)$ scales as $\exp(-n/t)$ when $n\gg t$ and crossovers to $(t/n)^2$ when $n\ll t$, the same as in \eqref{eq:heuristicconclusion}, and also in our discrete time model.  

Notice that as shown in Fig.~\ref{fig:sum_f},  $\sum f_{n\geq 1}$ also quickly saturates to a constant.  This demonstrates that while the early time dynamics of our master equation is not exact, the late time physics is quantitatively consistent with the microscopic constraint (\ref{eq:projection}).  

\section{Discussion}
\label{sec:discussion}
In this paper, we construct a one dimensional non-unitary free fermion circuit model with non-trivial steady state. Through extensive numerical calculation, we demonstrate that this model has emergent criticality and has two dimensional conformal symmetry.  The critical behavior observed in our circuit model is very robust and is insensitive to the parameters of the model. To understand this universal dynamics, we provide an interpretation in terms of the fast spreading of $C$ matrix in real space, which can be estimated by a classical non-linear master equation which also exhibits emergent conformal invariance. We expect other non-unitary quantum dynamical systems could also exhibit this ``quantum self-organized criticality", which arises without fine-tuning any parameters. We note in passing that ``quantum critical phases" that appear in generic regions of phase space (without fine tuning) naturally arise in finite density holographic matter \cite{Faulkner:2009wj,Iqbal:2011in,Hartnoll_2016apf}; however, these quantum critical phases are not stabilized by non-unitary or through dynamics.

The coefficient in front of $\log L_A$ entanglement entropy, interpreted as the effective central charge, is sensitive to various system parameters, which suggests that this CFT is not unitary or rational.  Non-unitary CFTs have arisen previously in many studies of random systems \cite{Chalker_1988,Cardy_1997,Cardy_2000,Gurarie_2002,Cardy_2013}. Since analytically solving such random systems is quite challenging, our model provides a new and simple example that could be a starting point to explore non-unitary CFTs in two dimensions.

The mutual information bewteen two disjoint intervals is a function of cross ratio and can provide more information about the structure of the CFT beyond the effective central charge \cite{Calabrese_2004,Calabrese_2009,Calabrese_MI_2009,Calabrese_MI_2011}. Previous research in the rational CFT shows in the small $\eta$ expansion, the mutual information encodes rich information of the scaling dimensions of the operator contents and the operator product expansion (OPE) coefficients \cite{Calabrese_MI_2011}. We expect such general principle also works in this non-unitary CFT. Notice in our model, $I(\eta)\sim \eta$ in the $\eta\to 0$ limit, indicating that in the OPE of the twist field, the lowest scaling dimension of the allowed operator is 1. It would be interesting to have a better understanding of this scaling dimension and examine how universal this result is in the future.

We expect that our model can break down in the presence of interactions between fermions.  Previous studies of random quantum dynamics with weak measurements found a phase transition between volume-law and area-law entangled phases \cite{Li_2019, Szyniszewski_2019}.  While our model is \textit{not} simply a proxy for weak measurements, it is possible that emergent criticality survives until a critical non-zero interaction strength. It is interesting to understand better the nature of this phase transition at finite interaction strength; we leave this problem for future study.

Last but not the least, we briefly discuss the realization in experiments. The non-unitary circuit presented in this paper can be thought of modelling some kind of stochastic non-hermitian Hamiltonian dynamics, which can be realized in an open quantum system under continuous measurements. Thus formulated, a possible experimental realization of the non-unitary circuit faces the same challenges posed for unitary-measurement circuits \cite{Skinner_2019, Li_2018, Chan_2019, gullans2019dynamical, gullans2019scalable, zabalo2019critical, choi2019quantum, Tang_Zhu_2020, Li_2019, Szyniszewski_2019, zhang2020nonuniversal, goto2020measurementinduced, jian2019measurementinduced, bao2019theory,fan2020selforganized}: in order to measure any entanglement measures of the final state of circuit evolution, one needs to prepare several/many copies of the same wavefunction, which requires post-selection on full-counting trajectories from an ensemble of exponentially many trajectories; therefore, the circuit needs to be run exponentially many times.
It might be possible to reduce such overheads in non-unitary circuit models.


\acknowledgements
We acknowledge useful discussions with Xiangyu Cao, Ying Ran, Shinsei Ryu and Tianci Zhou, and thank Paul Krapivsky for alerting us to \cite{KRAPIVSKY1993157}. XC acknowledges support from the DARPA DRINQS program. YL and MPAF acknowledge support from the Heising-Simons Foundation.

\appendix 
\section{Time evolution of $C$ matrix}
\label{app:C_algorithm}
In this appendix, we discuss two methods to compute the time evolution of correlation function $C$ matrix. Both methods give the same results. In our paper, we use the first method to numerically compute $C$ matrix. On the other hand, the equation of motion derived in the second method will be used to derive the master equation in Appendix \ref{app:mastereqn}.

\emph{Method 1.}  For a Hermitian Hamiltonian $H=\sum_{ij}H_{ij}c^\dag_ic_j$, if we take an initial product state 
\begin{align}
|\psi_0\rangle=\prod_{k=1}^N c^\dag_k|0\rangle,
\end{align}
under unitary time evolution $U=\exp(-iHt)$, we have
\begin{align}
|\psi_1\rangle=\prod_{k=1}^N c^\dag_k(t)|0\rangle,
\end{align}
where the Heisenberg operator $c_k^\dag(t)\equiv U^\dag c^\dag_k U=\sum_{j} U^\dag_{jk} c_j^\dag$.
The unitary evolution of these $c^\dag_i(t)$ with $1\leq i \leq N$ can be characterized by a $W$ matrix, which is the first $N$ columns of $U$ matrix and has dimension $L\times N$. The $C$ matrix can be evaluated as following,
\begin{align}
C_{ij}=\left(WW^\dag\right)^T_{ij}.
\label{eq:C_mat}
\end{align}

Similarly, for the wave function under imaginary time evolution $V=\exp(-H \beta)$, 
we can also define a $W$ matrix to characterize the wave function. We first choose the first $N$ columns of $V$ matrix and use them to construct an orthonormal basis from it, which forms $W$ matrix. We then use Eq.\eqref{eq:C_mat} to compute $C$ matrix. The physics behind this algorithm is very simple: Under imaginary time evolution,
\begin{align}
|\psi_2\rangle\sim \prod_{k=1}^N c_k^\dag(\beta)|0\rangle,
\end{align}
where $c_k^\dag(\beta)=Vc_k^\dag V^{-1}=\sum_j V_{jk}c_j^\dag$. However, since $V$ is not a unitary matrix, $\{c^\dag_i(\beta), c_j(\beta)\}\neq 0$ when $i\neq j$. We can construct a new canonical basis $f_k^\dag$ from $c^\dag_k(\beta)$ which satisfies the anti-commutation relations. In this basis, the wave function can be simply written as  
\begin{align}
|\psi_2\rangle=\prod_{k=1}^N f^\dag_k|0\rangle.
\end{align}

\emph{Method 2.}  In the second method, we directly compute the equation of motion for $C$ matrix. Under unitary time evolution, 
\begin{align}
C_{ij}(t)\equiv \langle \psi_1|c^\dag_ic_j|\psi_1\rangle=\langle \psi_0| U^\dag c^\dag_ic_j U|\psi_0\rangle.
\end{align}
By taking derivative of $C_{ij}(t)$,we have 
\begin{align}
\frac{dC_{ij}}{dt}&=\sum_{kl}iH_{kl}  \left [\langle c^\dag_kc_lc^\dag_i c_j\rangle-\langle c_i^\dag c_j c_k^\dag c_l\rangle\right] \nonumber\\
&=\sum_k i H_{ki}C_{kj}-\sum_{l} i C_{il} H_{jl}.
\end{align}
The second equation is obtained by using Wick theorem. Therefore we have
\begin{align}
\frac{d C}{dt}=i [H^T,C] \longrightarrow C(t)=e^{i H^T t} C(0) e^{-i H^T t}.
\label{eq:unitary_C}
\end{align} 
Under imaginary time evolution, we have
\begin{align}
C_{ij}=\frac{\langle \psi_0| V c_i^\dag c_j V|\psi_0\rangle}{\langle \psi_0| V V |\psi_0\rangle}.
\end{align}
This leads to
\begin{align}
\frac{d C_{ij}}{d\beta}&=\sum_{kl}H_{kl}\left[ -\langle c^\dag_kc_lc^\dag_ic_j \rangle-\langle c^\dag_ic_jc^\dag_kc_l \rangle+2\langle c^\dag_ic_j\rangle\langle c^\dag_kc_l\rangle\right]\nonumber\\
&=-\sum_{kl}H_{kl}\left[ C_{kj}(\delta_{li}-C_{il})+C_{il}(\delta_{jk}-C_{kj}) \right].
\end{align}
Therefore, we have 
\begin{align}
\frac{d C}{d\beta}=-\{ H^T,C \}+2CH^TC.
\label{eq:imaginary_C}
\end{align}
The nonlinear term $CH^T C$ is very important and is responsible for the interesting dynamics observed in our circuit model. 

\section{Numerical results for variants of our model}
\label{app:more_num}
\begin{figure*}[hbt]
\centering
\subfigure[]{
  \label{fig:Corr_p1_0_p2_05}
  \includegraphics[width=.65\columnwidth]{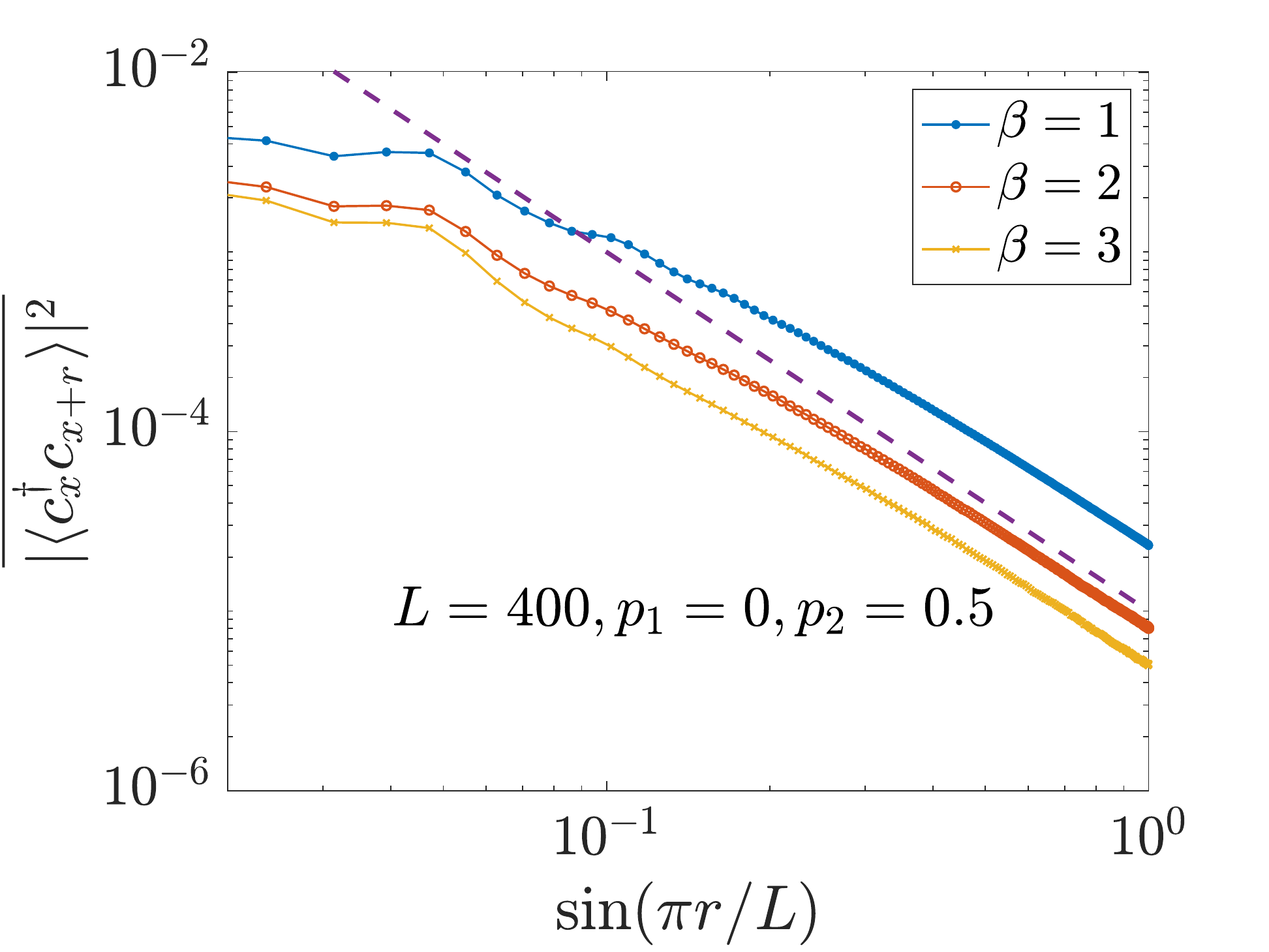}
}
\subfigure[]{
  \label{fig:S1_LA_p1_0}
  \includegraphics[width=.65\columnwidth]{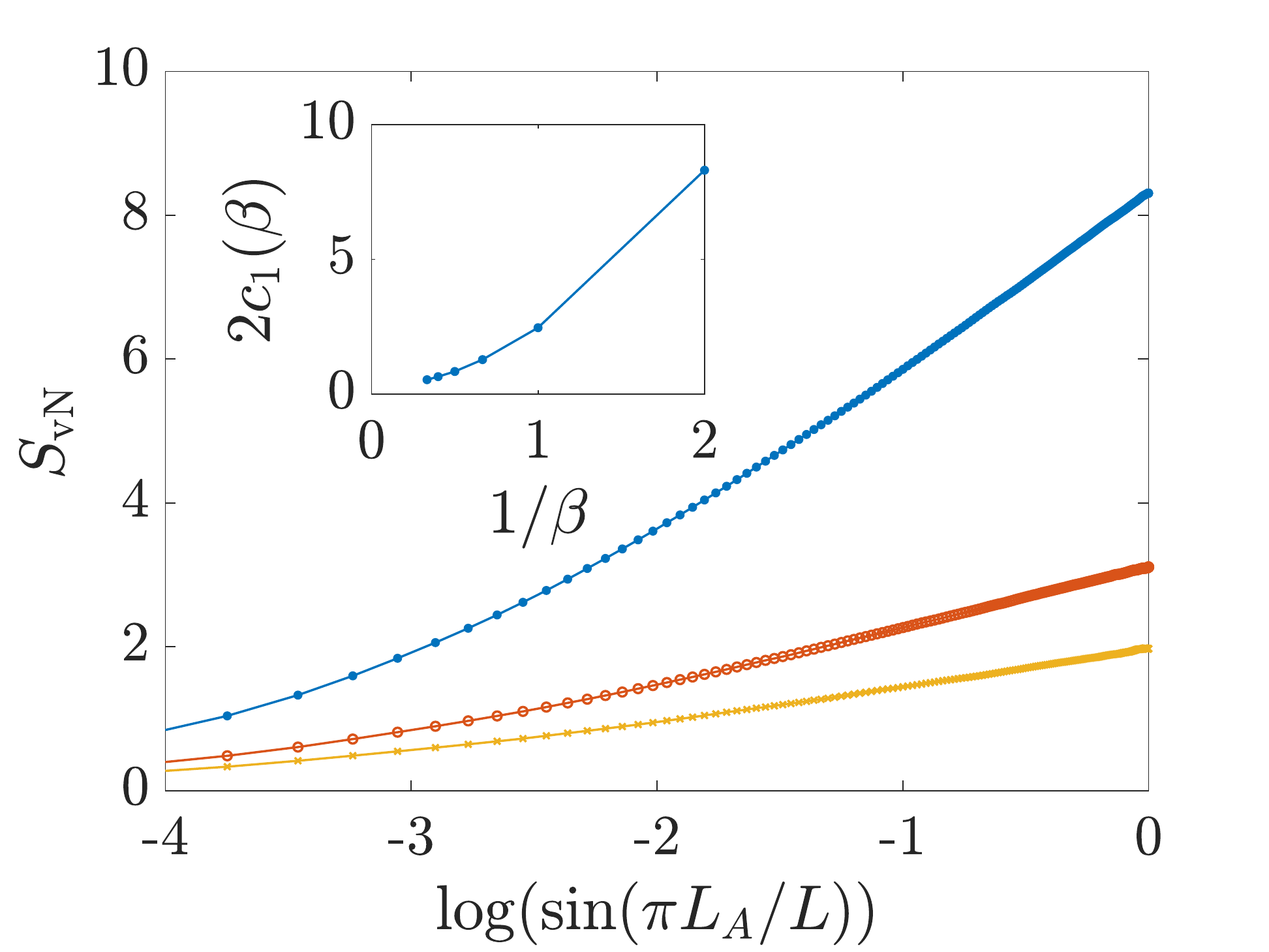}
}
\subfigure[]{
  \label{fig:MI_eta_p1_0_p2_05}
 \includegraphics[width=.65\columnwidth]{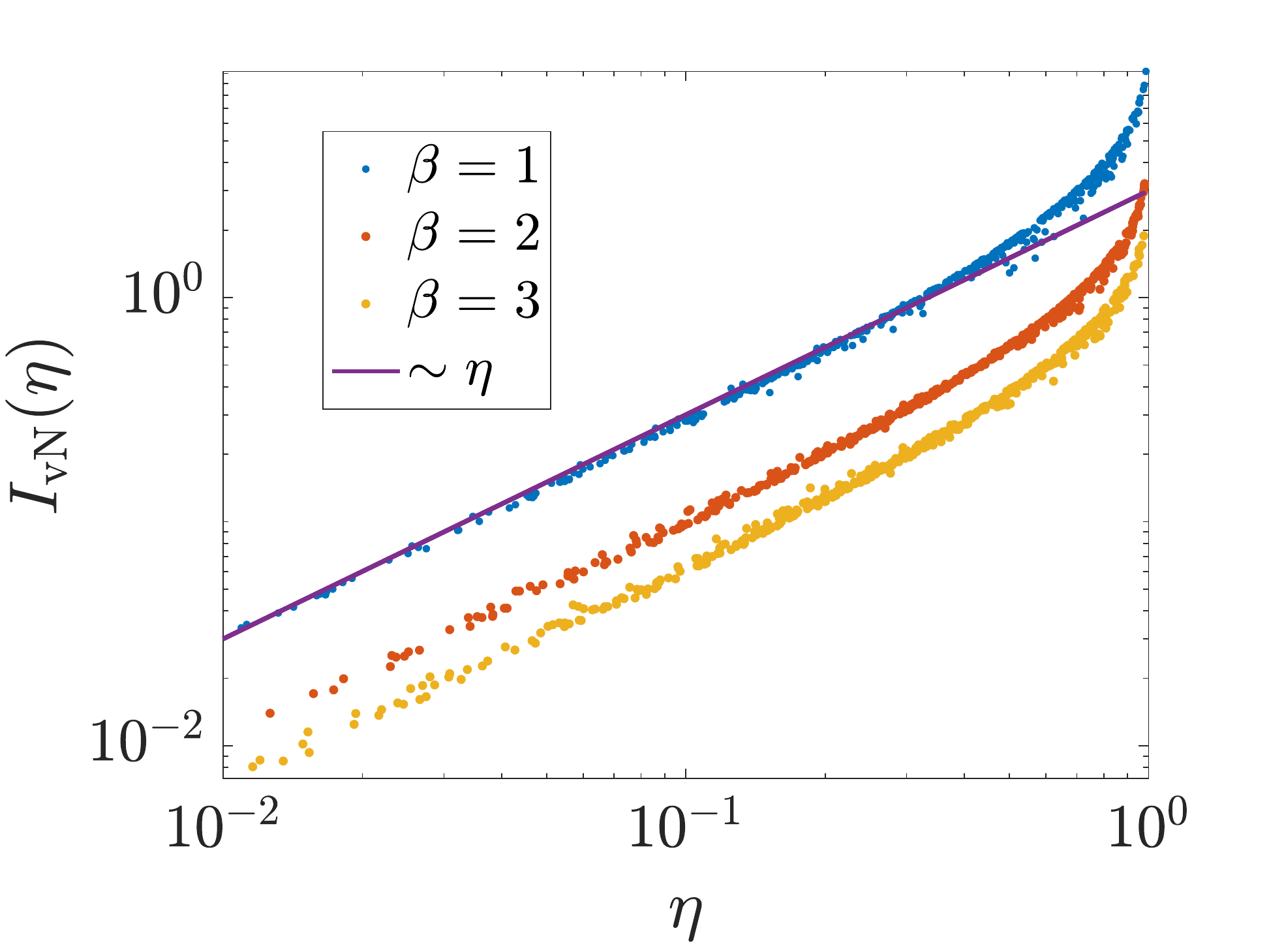}
}
\caption{The numerical results of the steady state for various $\beta$ at $p_1=0$ and $p_2=0.5$ with $L=400$ at $1/2$ filling. Here we take periodic boundary condition. (a) Squared correlation function on the log-log scale. The slope of the curves is 2 and is the same   as the dashed line which scales as $1/(\sin(\pi r/L))^2$. (b) von Neumann entanglement entropy $S_{\rm vN}$ vs $\log(\sin(\pi L_A/L))$ on the linear scale. The curves have the same labelling as in (a). The coefficient $2c_1(\beta)$ vs $1/\beta$ is shown in the inset. (c) Data collapse for the mutual information $I_1$ as a function of the cross ratio $\eta$ on the log-log scale. The locations of $x_i$ are chosen randomly on the circle with the constraint $|x_i-x_j|>3$.}
\label{fig:p1_0_p2_05}	
\end{figure*}

\begin{figure*}[hbt]
\centering
\subfigure[]{
  \label{fig:Corr_p1_03_p2_03}
  \includegraphics[width=.65\columnwidth]{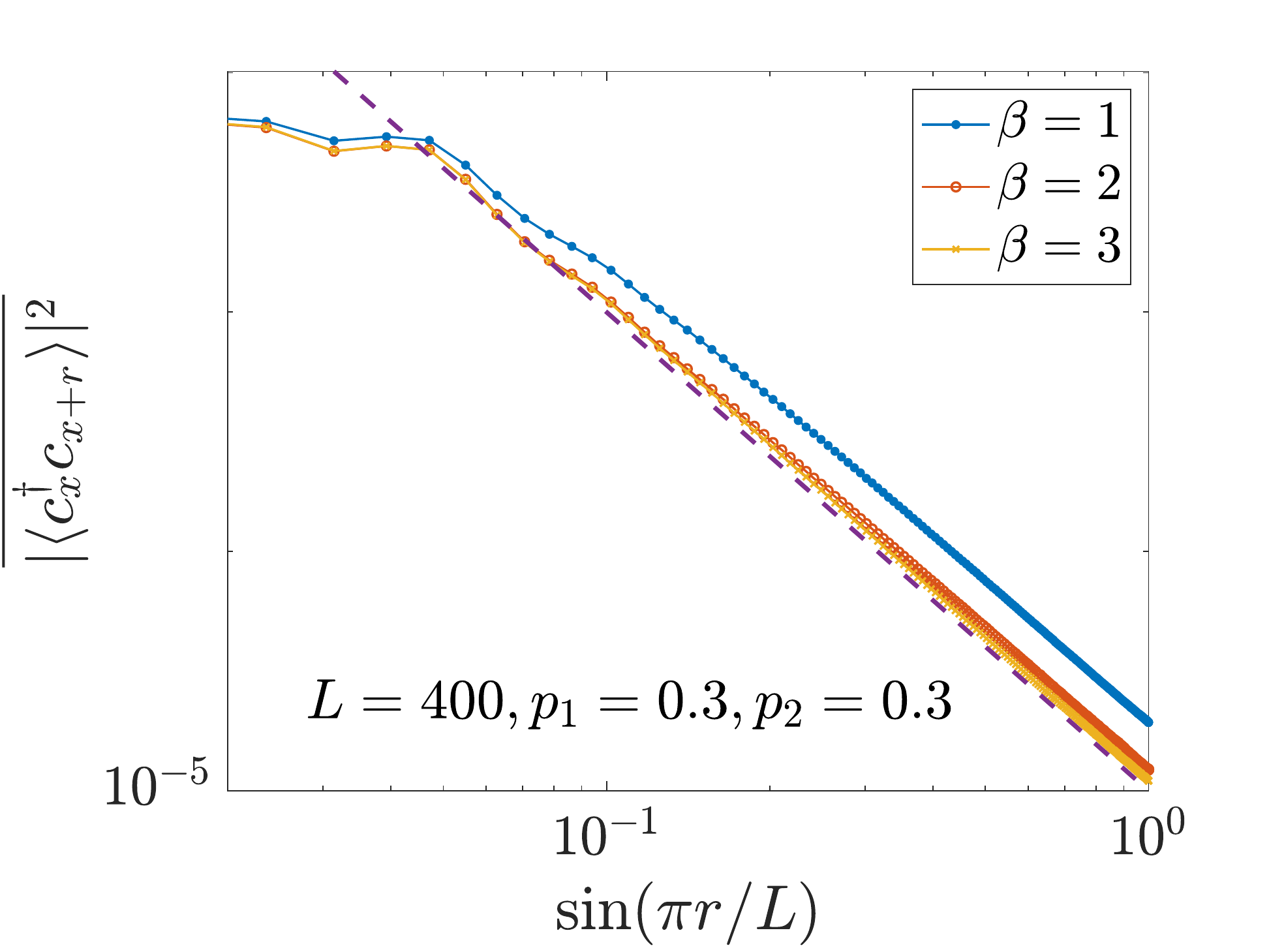}
}
\subfigure[]{
  \label{fig:S1_LA_p1_03}
  \includegraphics[width=.65\columnwidth]{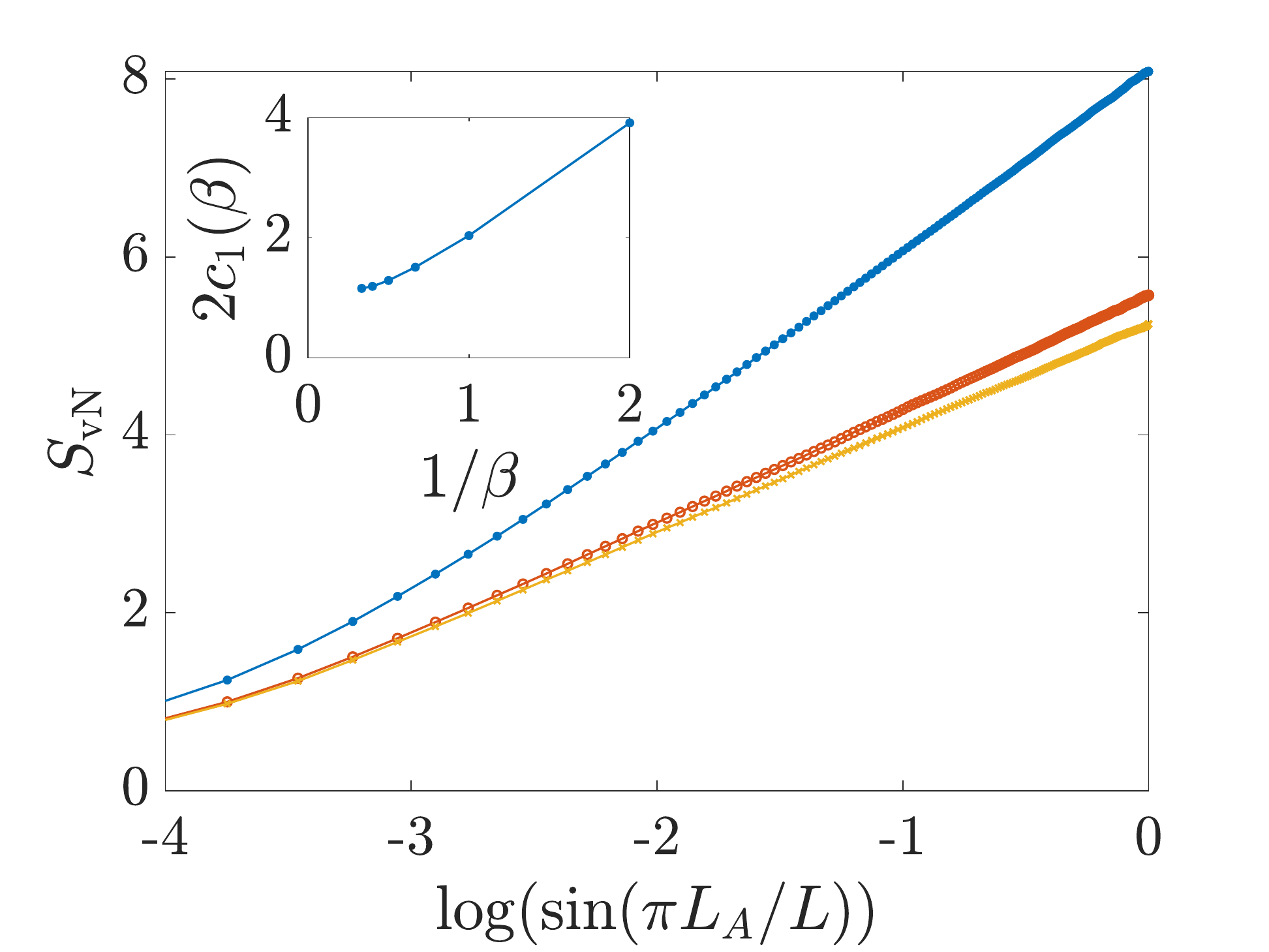}
}
\subfigure[]{
  \label{fig:MI_eta_p1_03_p2_03}
 \includegraphics[width=.65\columnwidth]{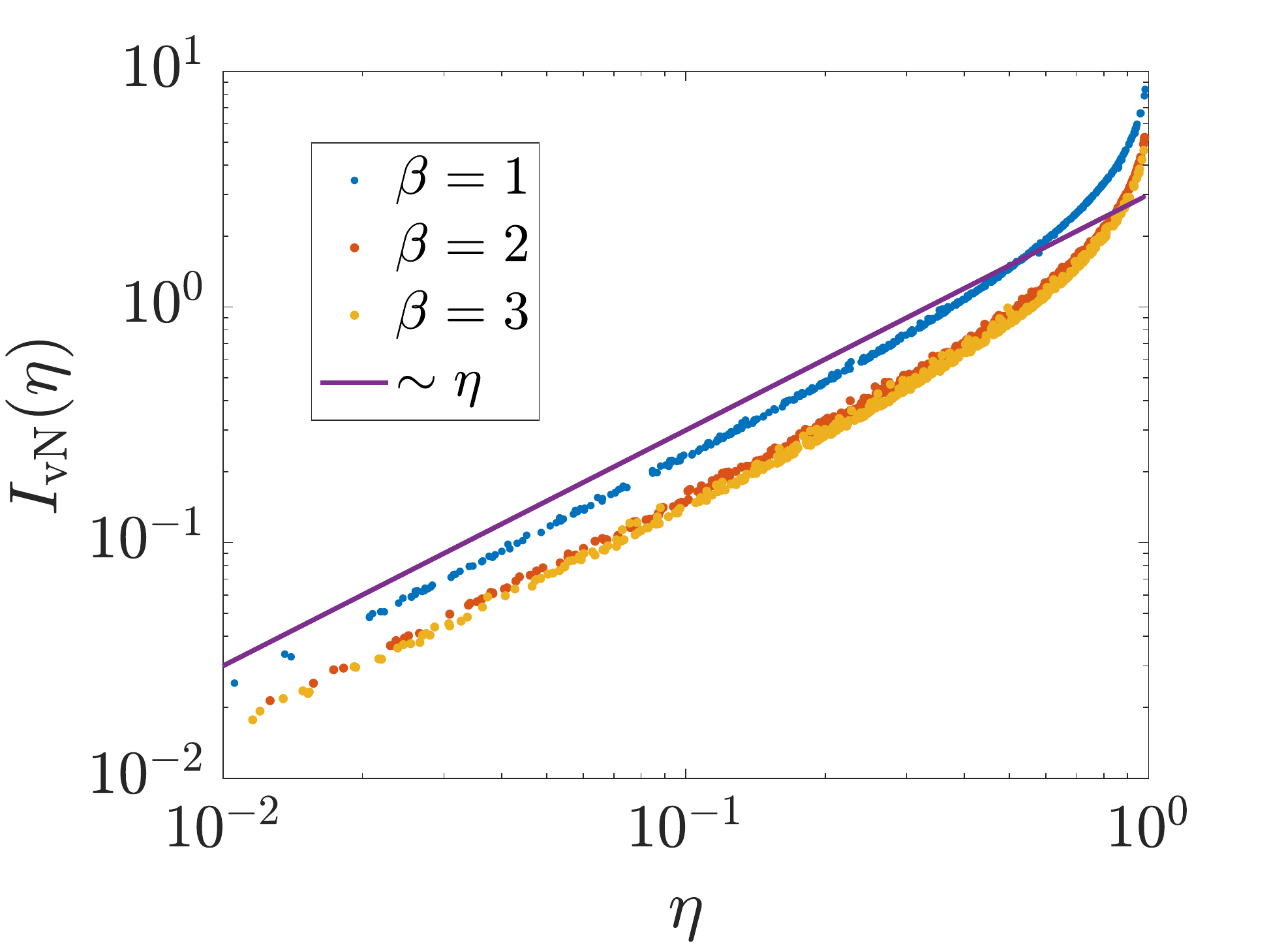}
}
\caption{The numerical results of the steady state for various $\beta$ at $p_1=p_2=0.3$ with $L=400$ at $1/2$ filling. Here we take periodic boundary condition. (a) Squared correlation function on the log-log scale. The slope of the curves is 2 and is the same   as the dashed line which scales as $1/(\sin(\pi r/L))^2$. (b) von Neumann entanglement entropy $S_{\rm vN}$ vs $\log(\sin(\pi L_A/L))$ on the linear scale. The curves have the same labelling as in (a). The coefficient $2c_1(\beta)$ vs $1/\beta$ is shown in the inset. (c) Data collapse for the mutual information $I_1$ as a function of the cross ratio $\eta$ on the log-log scale. The locations of $x_i$ are chosen randomly on the circle with the constraint $|x_i-x_j|>3$.}
\label{fig:p1_03_p2_03}	
\end{figure*}

We consider the model described in Eq.\eqref{eq:non_unitary} and take other values for $p_1$ and $p_2$. As shown in Fig.~\ref{fig:p1_0_p2_05} and Fig.~\ref{fig:p1_03_p2_03}, the steady state shows the same scaling behavior while the coefficient $c_1$ in front of $\log L_A$ scaling is model dependent. In particular, in Fig.~\ref{fig:p1_0_p2_05}, we take $p_1=0$ so that the randomness in the unitary evolution is turned off. These results strongly indicate that this critical behavior is robust and is not sensitive to these parameters in the model. 

\begin{figure*}[hbt]
\centering
\subfigure[]{
  \label{fig:S_p1_0_collapse_beta}
  \includegraphics[width=.8\columnwidth]{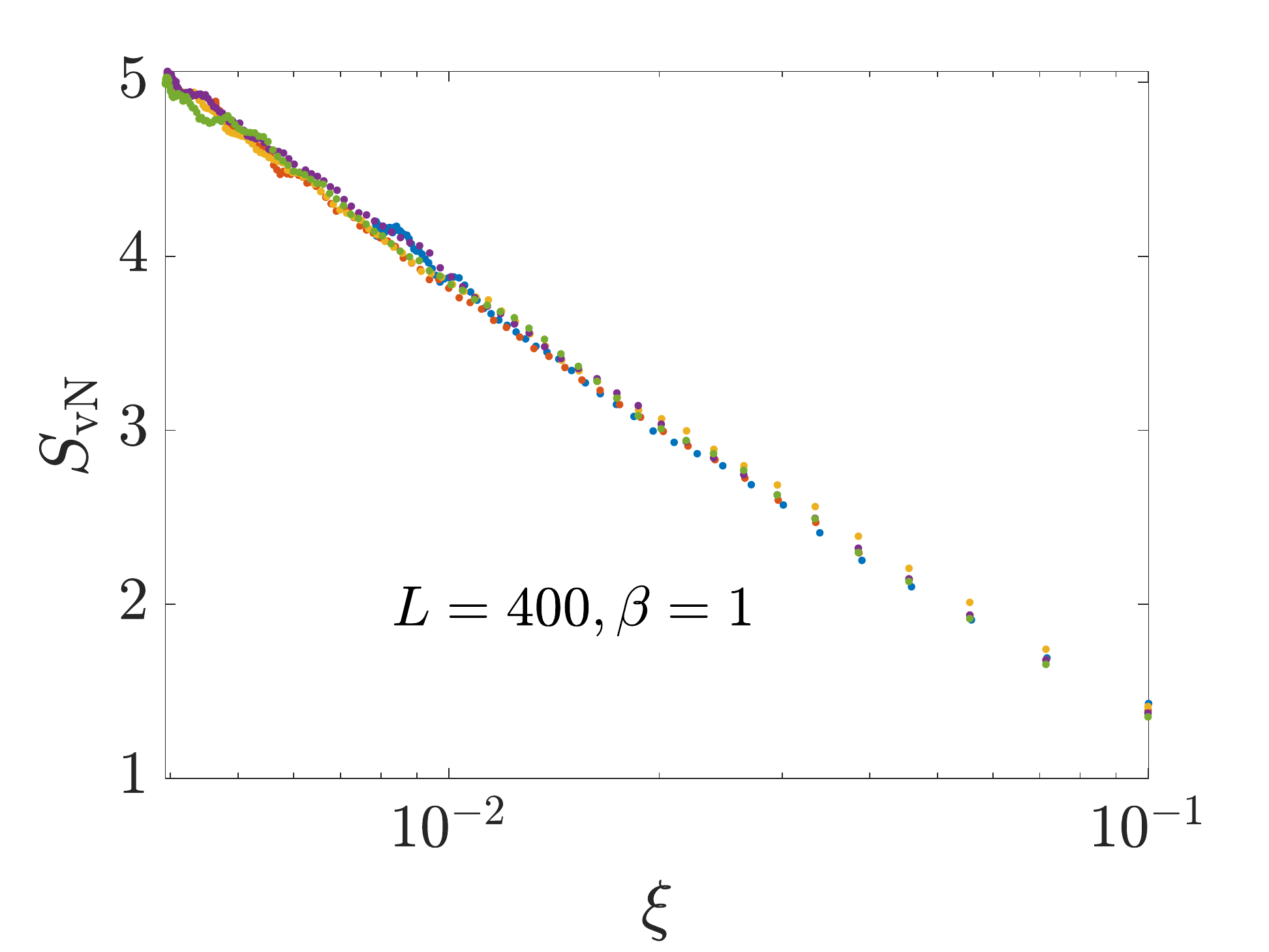}
}
\subfigure[]{
  \label{fig:MI_eta_p1_0_p2_05}
  \includegraphics[width=.8\columnwidth]{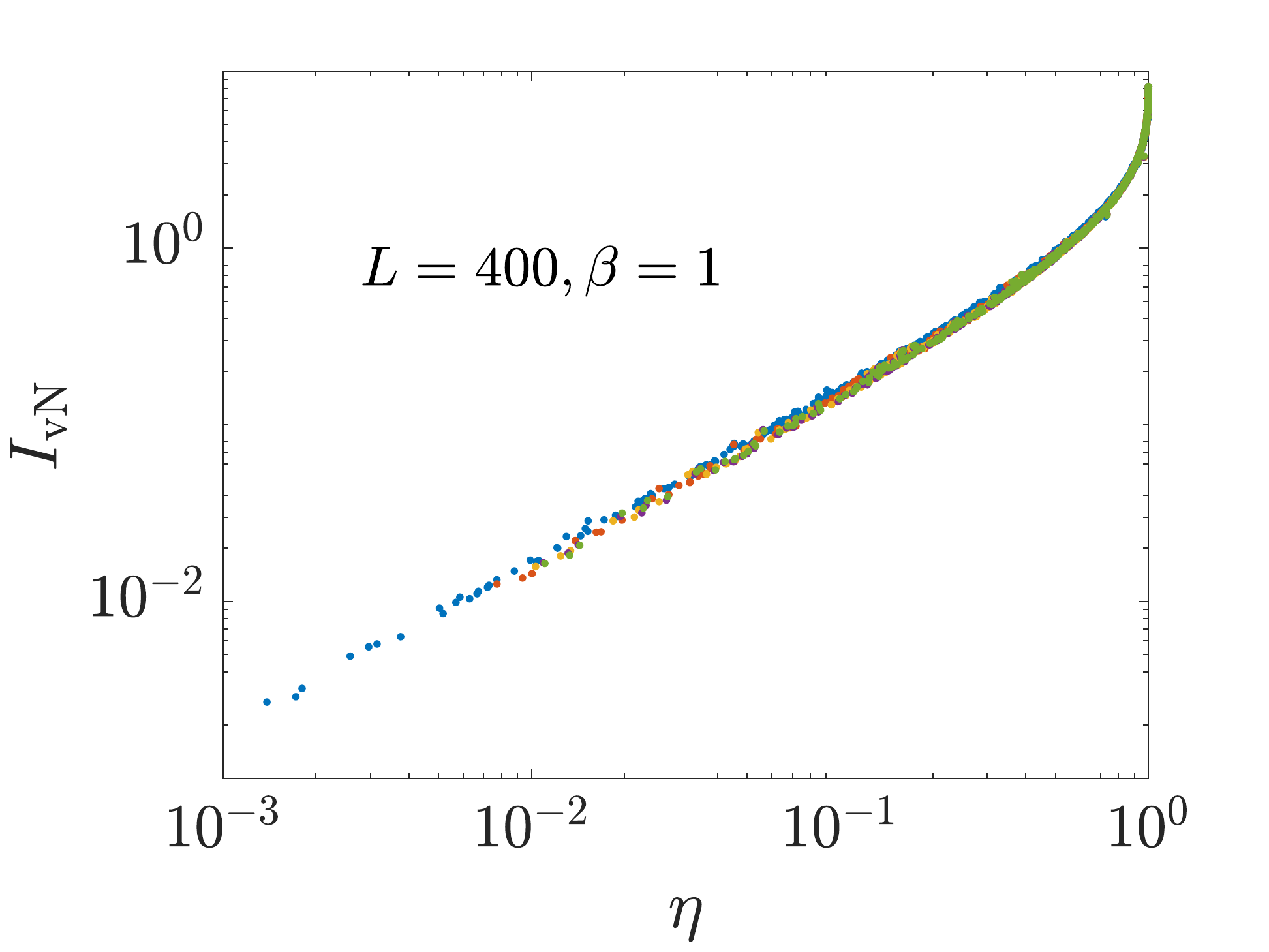}
}
\caption{The data collapse for the dynamics of entanglement entropy in (a) and mutual information in (b) in the regime $T\in[20, 100]$.  Here  we consider open boundary condition with $p_1=0$ and $p_2=0.5$.}
\label{fig:MI_p1_0_p2_05}	
\end{figure*}

\begin{figure*}[hbt]
\centering
\subfigure[]{
  \label{fig:S_p1_0_collapse_beta}
  \includegraphics[width=.8\columnwidth]{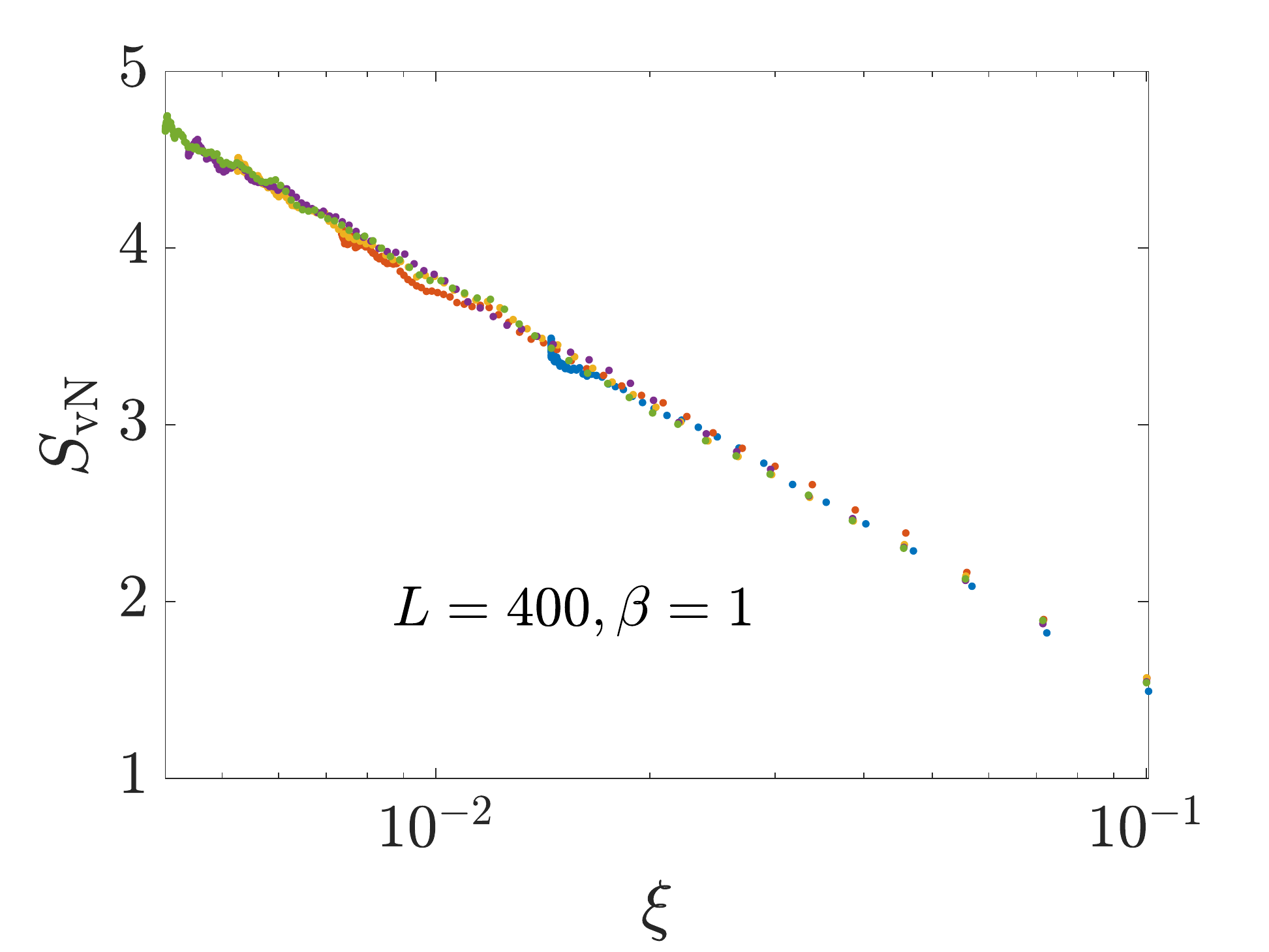}
}
\subfigure[]{
  \label{fig:MI_eta_p1_0_p2_05}
  \includegraphics[width=.8\columnwidth]{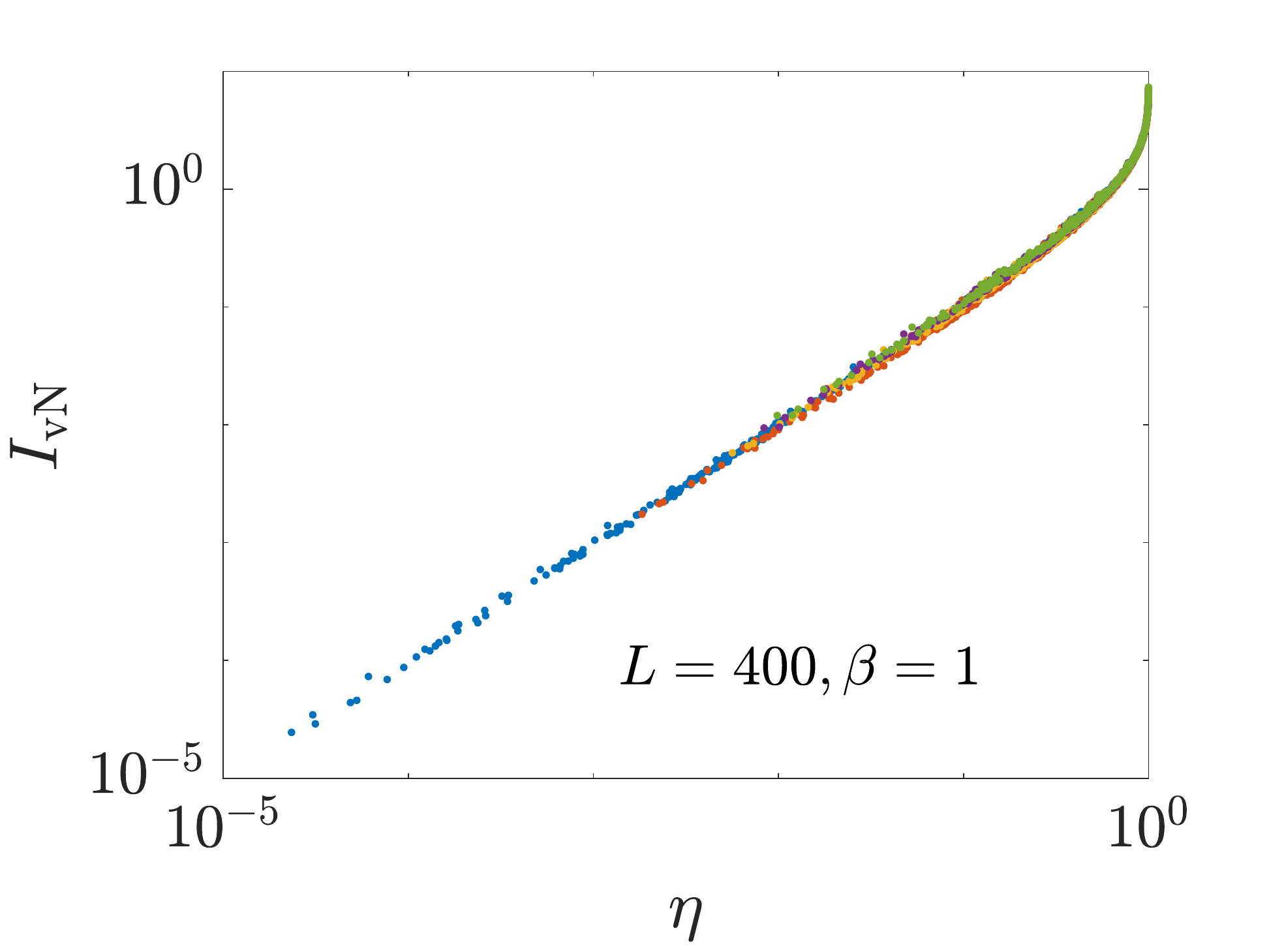}
}
\caption{The data collapse for the dynamics of entanglement entropy in (a) and mutual information in (b) in the regime $T\in[20, 100]$.  Here  we consider open boundary condition with $p_1=p_2=0.3$.}
\label{fig:MI_p1_03_p2_03}	
\end{figure*}

We further perform the data collapse for the dynamics of entanglement entropy and mutual information and present the results in Fig.~\ref{fig:MI_p1_0_p2_05} and  Fig.~\ref{fig:MI_p1_03_p2_03}. All the data points at different times collapse into a single curves. For mutual information, we observe that $I_{\rm vN}\sim \eta$ when $\eta$ is small and this is also true for other R\'enyi indices. 

\section{Absence of light cone}
\label{app:nolinearlightcone}
\begin{figure}[hbt]

\includegraphics[width=.65\columnwidth]{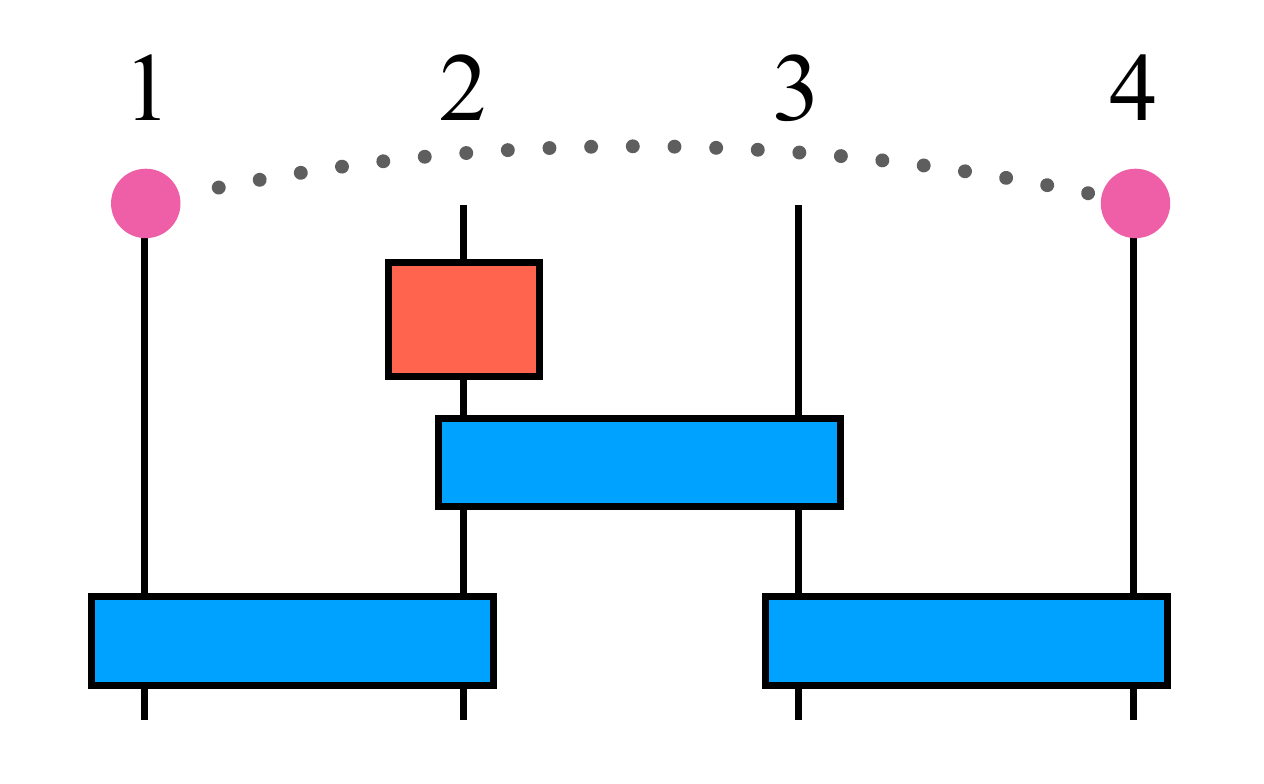}
\caption{The schematics for the non-unitary fermionic dynamics with 4 sites. The blue blocks $U_{ij}$ are two-qubit unitary gates defined in Eq.\eqref{eq:two_qubit_U}. The orange block denotes a non-unitary single qubit gate $\exp(-\beta n)$. This gate can induce nonzero correlation/entanglement between the first and fourth qubits.}

\label{fig:simple_cartoon}
\end{figure}

Here we give a simple cartoon picture (see Fig.~\ref{fig:simple_cartoon}) to explain how the non-unitary gates can break the linear light cone of quantum information dynamics.

Consider a fermionic system with 4 sites, initially prepared in the state \begin{equation}
    |\psi_0\rangle = |0101\rangle.
\end{equation}
Define the unitary gate \begin{align}
    U &= \frac{|01\rangle\langle 01| + |01\rangle\langle 10| - |10\rangle\langle 01| + |10\rangle\langle 10| }{\sqrt{2}} \notag \\
    &\;\;\; +  |00\rangle\langle 00| + |11\rangle\langle 11|
    \label{eq:two_qubit_U}
\end{align}
which entangles two sites if exactly one is empty.   
Let $U_{ij}$ denote this unitary gate acting on sites $i$ and $j$.  Consider first evolving the system with unitary dynamics:  \begin{align}
    |\psi_1\rangle &= U_{23}U_{34}U_{12}|\psi_0\rangle \notag \\
    &= \frac{|0101\rangle - |0011\rangle + |1010\rangle + |1100\rangle }{\sqrt{8}}\notag \\
    &\;\;\; -\frac{1}{2}|0110\rangle - \frac{1}{2}|1001\rangle.
\end{align}
Observe that the mixed state of fermions 1 and 4 is maximally mixed; hence the mutual information $I_{\psi_1}(1,4)=0$.

Now let us apply a non-unitary gate:  \begin{align}
    |\psi_2\rangle &= \frac{\mathrm{e}^{-\beta n_2}|\psi_1\rangle}{\sqrt{\langle \psi_1 |\mathrm{e}^{-2\beta n_2}|\psi_1\rangle}} \notag \\
    &\propto \frac{\mathrm{e}^{-\beta} |0101\rangle - |0011\rangle + |1010\rangle + \mathrm{e}^{-\beta}|1100\rangle }{\sqrt{8}}\notag \\
    &\;\;\; -\frac{\mathrm{e}^{-\beta}}{2}|0110\rangle - \frac{1}{2}|1001\rangle.
\end{align}
Observe that the factor $\mathrm{e}^{-\beta}$ weights different terms unequally.  The mixed state of sites 1 and 4 is no longer maximally mixed (nor that of site 1 alone).   When $\beta$ is small, we find the mutual information \begin{equation}
    I_{\psi_2}(1,4) = \frac{\beta^2}{16} + \cdots
\end{equation}
which is not vanishing, even though there is no chain of gates (running forward in time) which link site 1 to site 4.   The reason that this mutual information has arisen is the prefactors of the coefficients in $|\psi_1\rangle$ are finely tuned to ensure mutual information vanishes.  The non-unitary gate disrupts these tuned cancellations and thus generically spreads information instantaneously.

These two-qubit and one-qubit gates can be used to construct a many-body free fermion random circuit model.  We numerically compute this model and find the same critical behaviors with emergent conformal symmetry. 


\section{Cardy-Calabrese formalism}
\label{app:CC_formalism}

According to Cardy-Calabrese formalism\cite{Calabrese_2004,Calabrese_2009}, the entanglement entropy of the wave function
\begin{align}
|\psi(T)\rangle=\frac{e^{-T H_{\rm CFT}}}{\sqrt{Z}}|\psi_0\rangle
\end{align} 
for a $1+1$d CFT can be mapped to the correlation function for twist fields, i.e.,
\begin{align}
\mbox{Tr}\rho_A^n =\langle \mathcal{T}_n(z_1,\bar{z}_1)\ldots \mathcal{T}_n(z_m,\bar{z}_m) \rangle
\label{eq:Tr_rho_def}
\end{align}
where $\rho_A$ is the reduced density matrix for subsystem and $\mathcal{T}_n$ is the twist field operator with conformal dimension $\Delta_n=\overline{\Delta}_n=\frac{c}{24}(n-\frac{1}{n})$. The number of the twist field is determined by the geometry of the system and subsystem. These twist fields behave as the primary fields under conformal mapping and satisfy
\begin{align}
&\langle \mathcal{T}_n(z_1,\bar{z}_1)\ldots \mathcal{T}_n(z_m,\bar{z}_m) \rangle\nonumber\\
=&\left|\frac{\partial w}{\partial z}\right|_{z_1}^{2\Delta_n}\ldots \left|\frac{\partial w}{\partial z}\right|_{z_m}^{2\Delta_n}\langle \mathcal{T}_n(w_1,\bar{w}_1)\ldots \mathcal{T}_n(w_m,\bar{w}_m) \rangle.
\label{eq:conf_map}
\end{align}
Below we consider several simple geometries we are interested in this paper.

{\it Infinite plane}: Here we compute the steady state entanglement entropy for a single interval of an infinite long system. This corresponds to a simple infinite plane geometry with $L,T\to\infty$. According to Eq.\eqref{eq:Tr_rho_def}, we have 
\begin{align}
\mbox{Tr}\rho_A^n =\langle \mathcal{T}_n(z_1,\bar{z}_1)\ldots \mathcal{T}_n(z_m,\bar{z}_m) \rangle_{\mathcal{C}}=\left(\frac{1}{L_A}\right)^{4\Delta_n}
\end{align}
where $\langle\cdot\rangle$ is defined on the infinite plane and $L_A$ is the distance between two twist fields, which corresponds to the length of the subsystem. The Renyi entropy is equal to 
\begin{align}
S_n\equiv\frac{1}{1-n}\log \mbox{Tr}\rho_A^n=\frac{c}{6}\left(1+\frac{1}{n}\right)\log L_A.
\end{align}

\begin{figure*}[hbt]
\centering
\subfigure[]{
  \label{fig:cylinder}
  \includegraphics[width=.65\columnwidth]{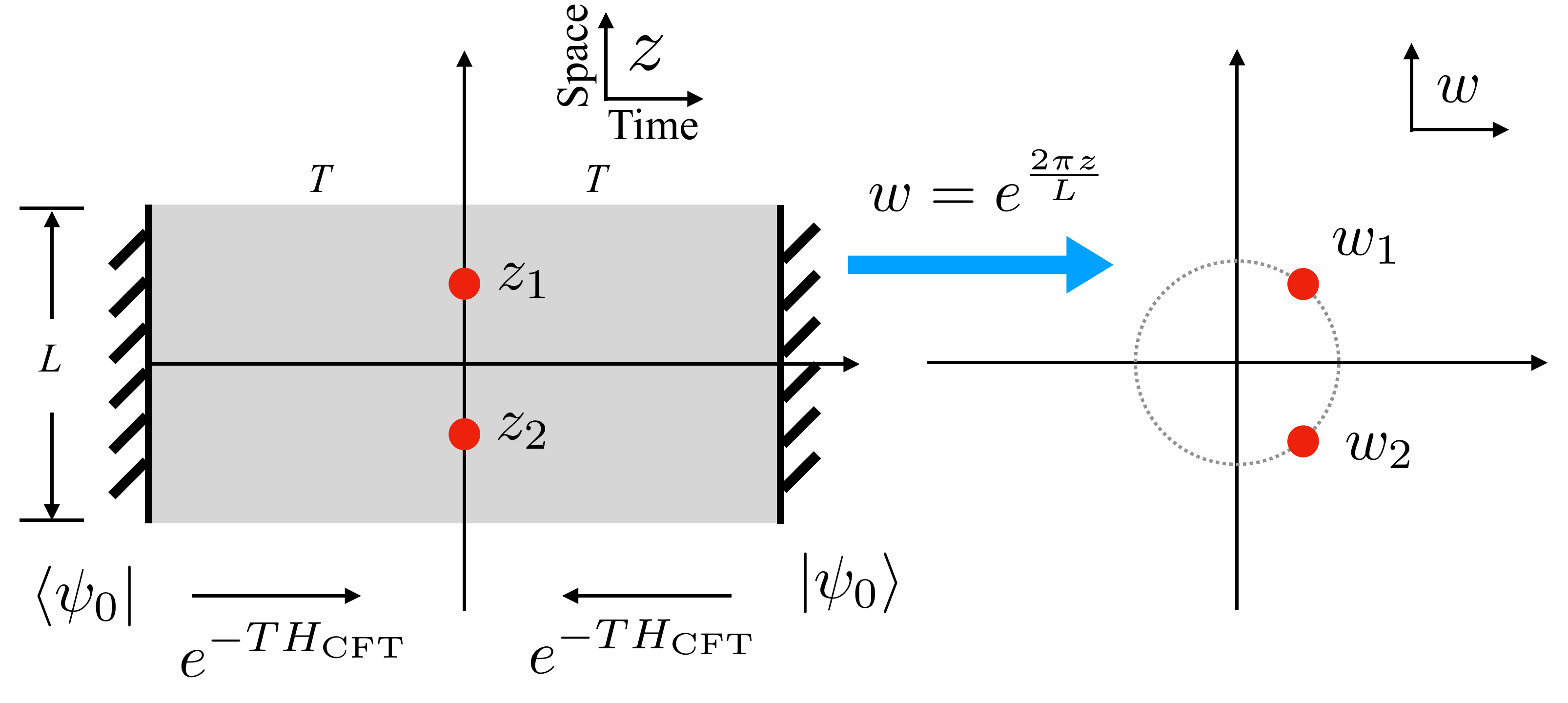}
}
\subfigure[]{
  \label{fig:strip}
  \includegraphics[width=.65\columnwidth]{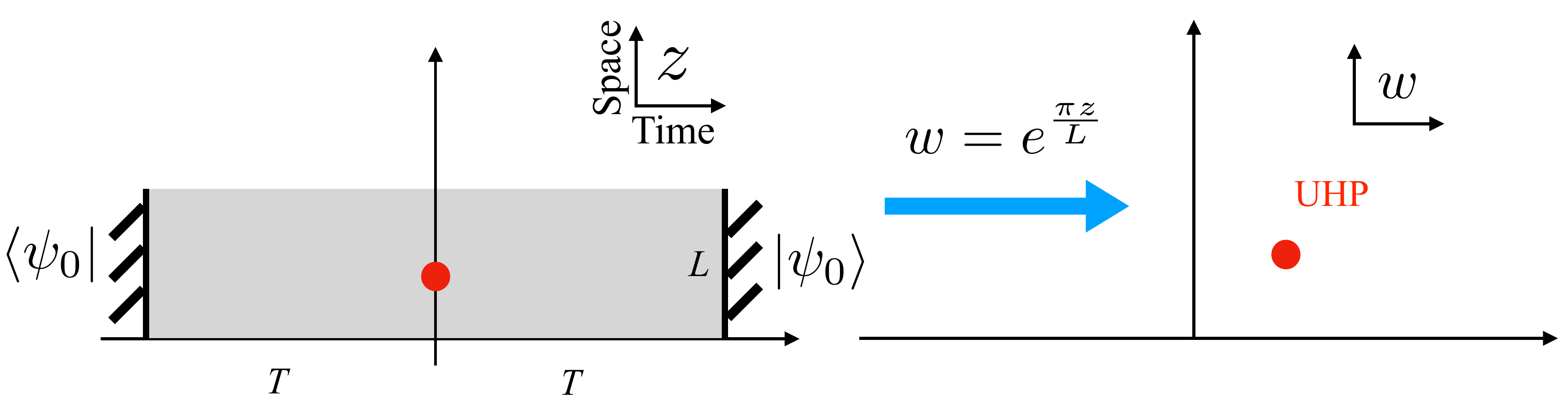}
}
\subfigure[]{
  \label{fig:semi_strip}
  \includegraphics[width=.65\columnwidth]{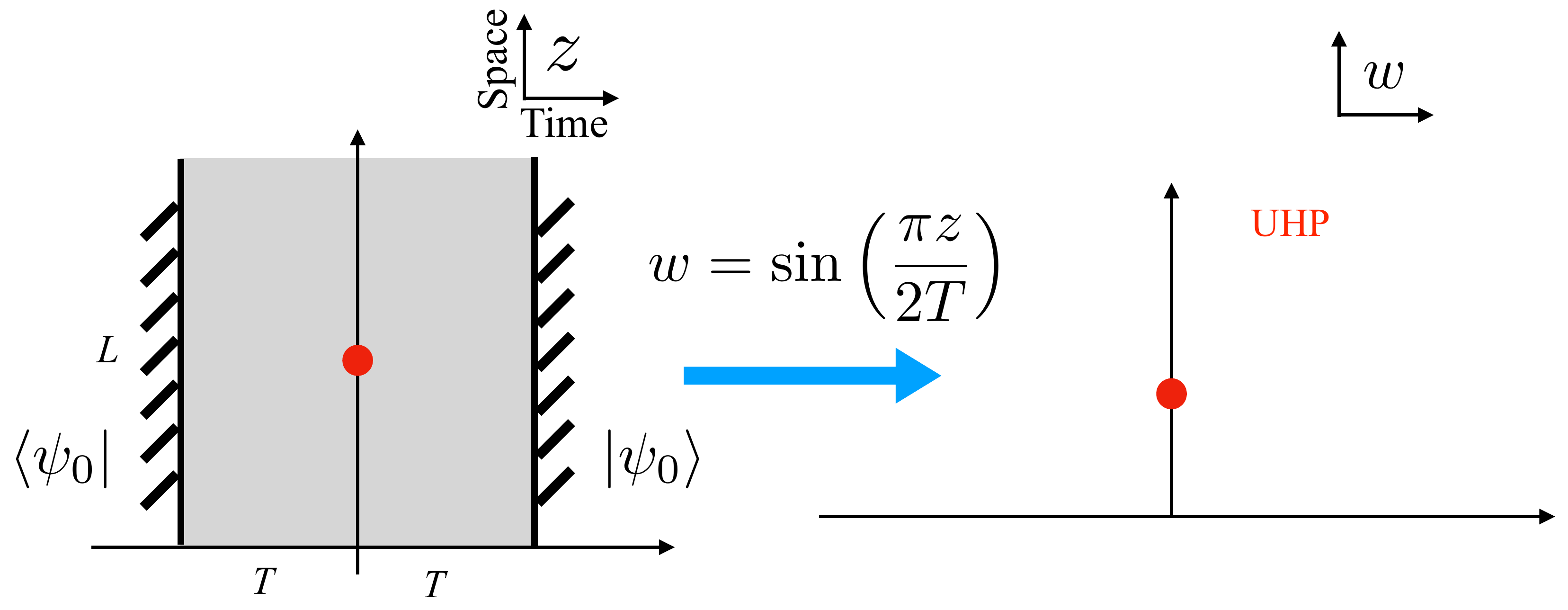}
}
\subfigure[]{
  \label{fig:rectangle}
  \includegraphics[width=.65\columnwidth]{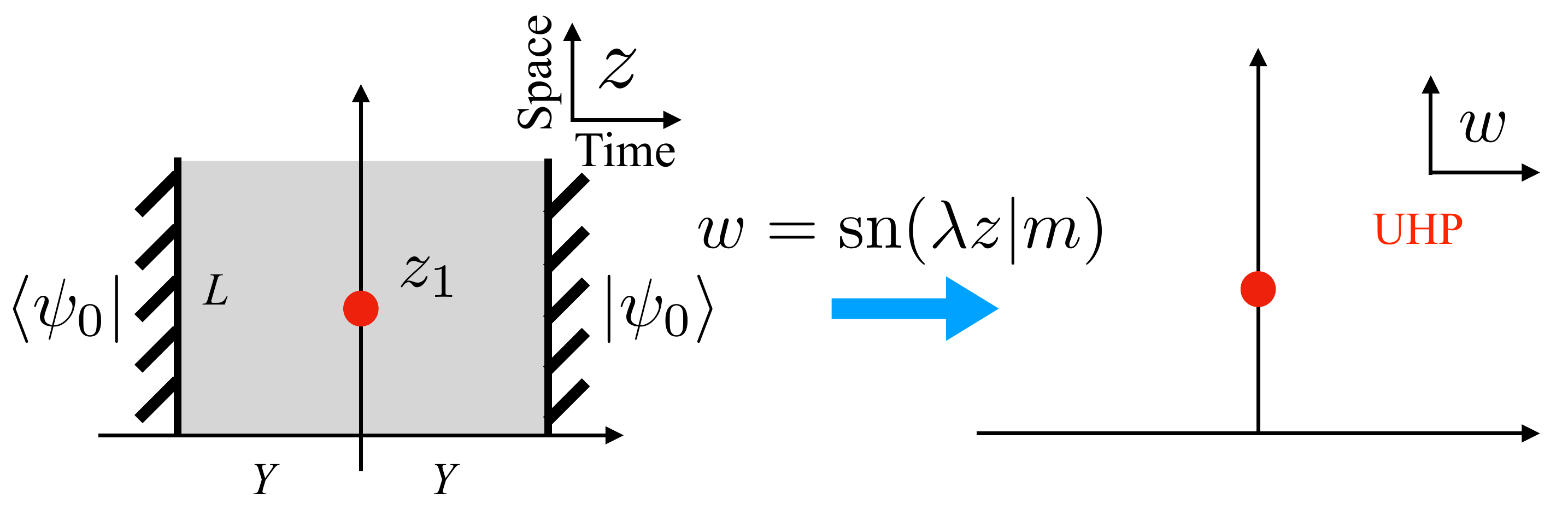}
}
\subfigure[]{
  \label{fig:sn_map}
  \includegraphics[width=.65\columnwidth]{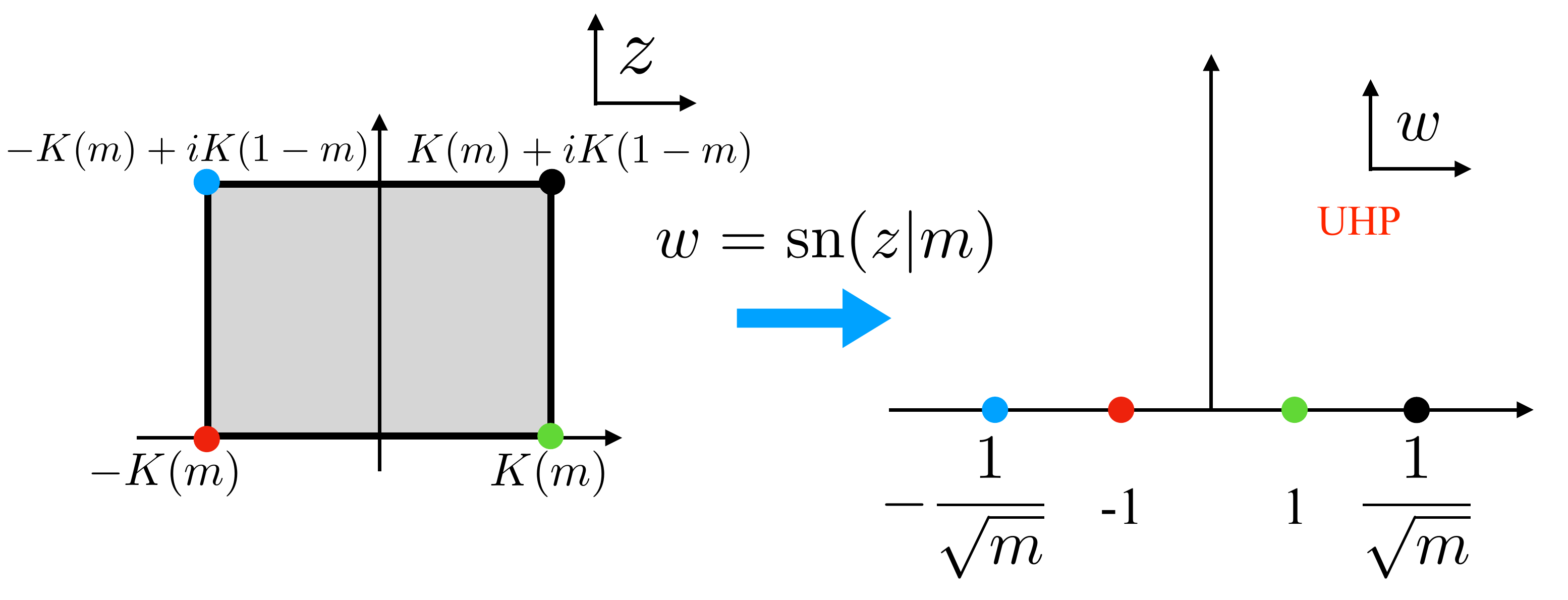}
}
\subfigure[]{
  \label{fig:rectangle_MI}
  \includegraphics[width=.65\columnwidth]{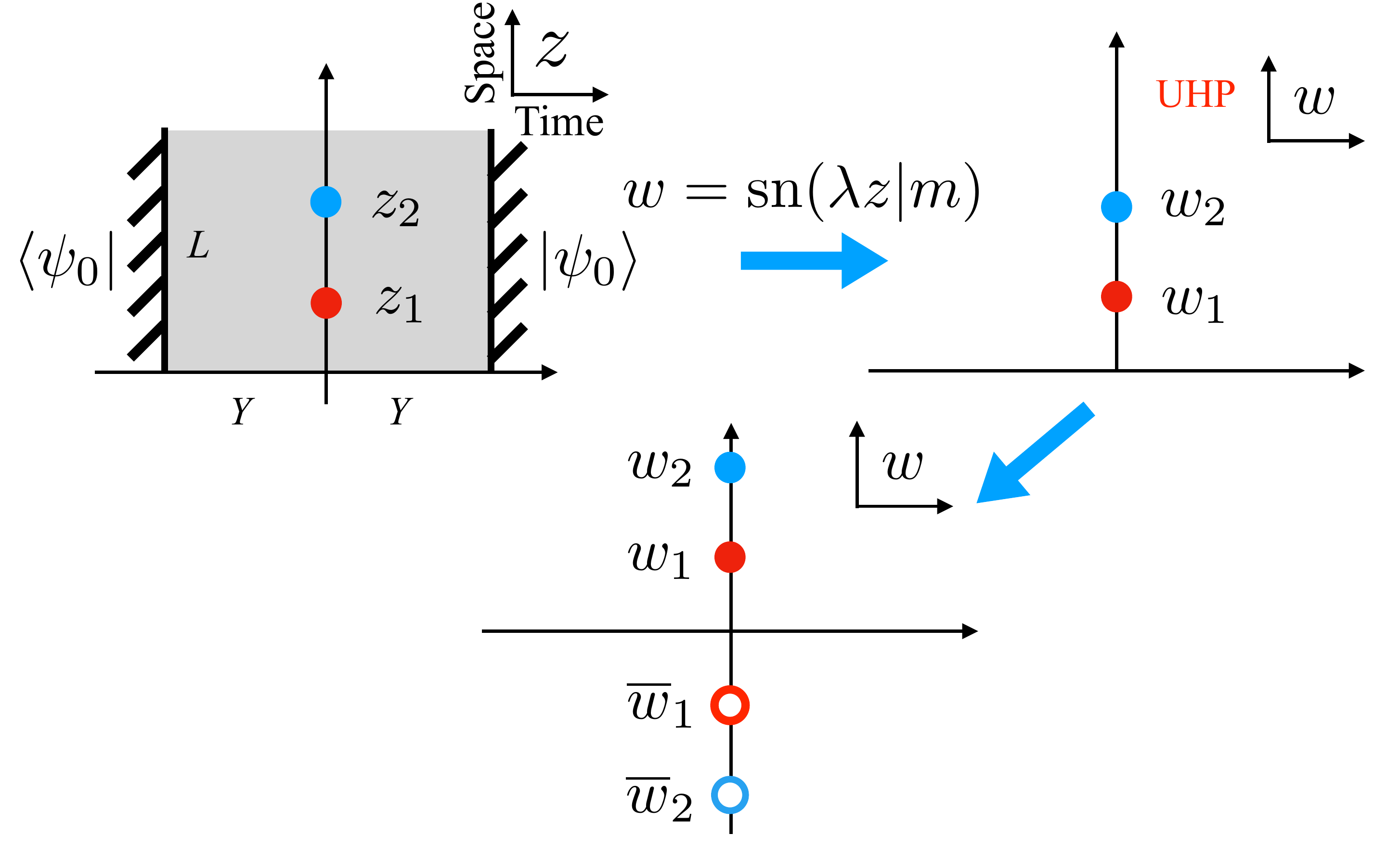}
}
\caption{(a) The conformal mapping from infinite cylinder ($T=\infty$) to a complex plane. In the left plot, the up and lower edges of the rectangle are glued together so that it is equivalent to a cylinder geometry. The left and right sides denote the short-range entangled initial state $|\psi_0\rangle$. The two twist fields are located at $z_1$ and $z_2$ with $z_1-z_2=iL_A$. After conformal mapping, $z_1$ and $z_2$ are mapped to $w_1$ and $w_2$ living on the unit circle around origin. (b) The conformal mapping from infinite strip ($L=\infty$) to the upper half plane (UHP). Here $z_1=iL_A$. (c) The conformal mapping from semi-infinite strip ($L=\infty$) to the upper half plane (UHP). Here $z_1=iL_A$. (d) The conformal mapping from a rectangle to UHP. Here $z_1=iL_A$. (e) The detail of the conformal mapping from a ``canonical" rectangle to UHP. The four corners of the canonical rectangle are mapped to four points on the real axis. (f) The procedure to evaluate the mutual information, where $z_1=iL_A$ and $z_2=i(L-L_B)$.}
\label{fig:conf_map}	
\end{figure*}

{\it Infinite cylinder}: Here we compute the steady state entanglement entropy for a single interval of one dimensional system with periodic boundary condition. This geometry corresponds to an infinite cylinder with the length $T=\infty$ and circumference $L$. As shown in Fig.~\ref{fig:cylinder}, the correlation function defined on the infinite cylinder can be computed by mapping it to a complex plane. By using Eq.\eqref{eq:conf_map}, we have
\begin{align}
\mbox{Tr}\rho_A^n=\langle T_n(z_1,\overline{z}_1) T_n(z_2,\overline{z}_2)\rangle=\left(\frac{2\pi}{L}\right)^{4\Delta_n}\frac{1}{(2\sin\frac{\pi L_A}{L})^{4\Delta_n}}
\end{align}
where $\langle\cdot\rangle$ is defined on the infinite cylinder. Thus we have
\begin{align}
S_n=\frac{c}{6}\left(1+\frac{1}{n}\right)\log\left[\frac{L}{\pi}\sin\left(\frac{\pi L_A}{L}\right)\right].
\label{eq:EE_pbc}
\end{align}
We further compute mutual information between two intervals A and B for the steady state with periodic boundary condition. This is related with a four point correlation function,
\begin{align}
&\frac{\mbox{Tr}\rho_{A\cup B}^n}{\mbox{Tr}\rho_{A}^n \mbox{Tr}\rho_{B}^n}=\frac{\langle \mathcal{T}_n(z_1,\overline{z}_1)\mathcal{T}_n(z_2,\overline{z}_2)\mathcal{T}_n(z_3,\overline{z}_3)\mathcal{T}_n(z_4,\overline{z}_4)\rangle}{\langle \mathcal{T}_n(z_1,\overline{z}_1) \mathcal{T}_n(z_2,\overline{z}_2)\rangle \langle \mathcal{T}_n(z_3,\overline{z}_3) \mathcal{T}_n(z_4,\overline{z}_4)\rangle}\nonumber\\
 &=\frac{\langle \mathcal{T}_n(w_1,\overline{w}_1)\mathcal{T}_n(w_2,\overline{w}_2)\mathcal{T}_n(w_3,\overline{w}_3)T_n(w_4,\overline{w}_4)\rangle}{\langle \mathcal{T}_n(w_1,\overline{w}_1) \mathcal{T}_n(w_2,\overline{w}_2)\rangle \langle \mathcal{T}_n(w_3,\overline{w}_3) \mathcal{T}_n(w_4,\overline{w}_4)\rangle}\nonumber\\
 &=F(\eta).
\end{align}
Therefore the mutual information is a function of cross ratio $\eta$, which is defined as
\begin{align}
\eta\equiv \frac{z_{12}z_{34}}{z_{13}z_{24}}
\end{align}
with $z_{ij}=\sin(\pi|z_i-z_j|/L)$.

{\it Infinite strip}: Here we compute the steady state entanglement entropy for a single interval of one dimensional system with open boundary. As shown in Fig.~\ref{fig:strip}, this requires us to compute the single point correlation function $\langle \mathcal{T}_n(z_1,\overline{z_1})\rangle$ defined on an infinite strip, which can be evaluated by mapping it to $\langle \mathcal{T}_n(w_1,\overline{w_1})\rangle_{\rm UHP}$ defined on the upper half plane (UHP). Notice that
\begin{align}
\langle \mathcal{T}_n(w_1,\overline{w}_1)\rangle_{\rm UHP}=\langle \mathcal{T}_n(w_1) \mathcal{T}_n(\overline{w}_1)\rangle_{\mathcal{C}}
\end{align}
where $\langle\cdot\rangle$ on the righthand side is defined on the complex plane and $\mathcal{T}_n(\overline{w}_1)$ is the image of $\mathcal{T}_n(w_1)$. We have
\begin{align}
\mbox{Tr}\rho_A^n=\left(\frac{\pi}{L}\right)^{2\Delta_n}\left(2\sin\frac{\pi L_A}{L}\right)^{-2\Delta_n},
\end{align}
which leads to
\begin{align}
S_n=\frac{c}{12}\left(1+\frac{1}{n}\right)\log\left[ \frac{2L}{\pi}\sin \left( \frac{\pi L_A}{L} \right) \right].
\end{align}

{\it Semi-infinite strip}: Here we compute the entanglement dynamics for a single interval of an infinite long system with open boundary. As shown in Fig.~\ref{fig:semi_strip}, this corresponds to evaluate the single point correlation function defined on the semi-infinite strip, which reads
\begin{align}
\mbox{Tr}\rho_A^n=\left(\frac{\pi}{4T}\right)^{2\Delta_n}\tanh^{-2\Delta_n}\left(\frac{\pi  L_A}{2T}\right).
\end{align}
In the limit $T\gg L_A$, we have 
\begin{align}
S_n=\frac{c}{12}\left(1+\frac{1}{n}\right)\log L_A;
\end{align}
In the limit $L_A\gg T$, we have
\begin{align}
S_n=\frac{c}{12}\left(1+\frac{1}{n}\right)\log T
\end{align}

{\it Rectangle geometry}: Finally we consider the entanglement dynamics for a single interval of finite system with open boundary. As shown in Fig.~\ref{fig:rectangle}, this corresponds to evaluate the single point correlation function defined on the rectangle. The conformal mapping from a rectangle to UHP is described by the Jacobi elliptic function ${\rm sn}(z|m)$. As shown in Fig.~\ref{fig:sn_map}, under this mapping, the four corners $\left[ -K(m)+iK(1-m),-K(m),K(m),K(m)+iK(1-m)  \right]$ of the rectangle map to the four points on the real axis. $K(m)$ is the complete elliptic integrable of the first kind
\begin{align}
K(m)\equiv\int_0^1 \frac{dt}{\sqrt{(1-t^2)(1-m t^2)}}.
\end{align}
The aspect ratio of the rectangle is $2K(m)/K(1-m)$.

In our numerical simulation, the rectangle has length $2T$ and height $L$. The aspect ratio is defined as 
\begin{align}
\tau=a\frac{2T}{L}=\frac{2Y}{L},
\label{eq:aspect_ratio}
\end{align}
where we introduce a parameter $a$ which is model dependent and rescales the time direction. Given a cross ratio $\tau$, we numerically find the corresponding $m$. We then map this $L\times 2Y$ rectangle to UHP with the following conformal mapping: 
\begin{align}
w(z)={\rm sn}(\lambda z|m),
\end{align}
where $\lambda=L/K(1-m)$.

Therefore we have
\begin{align}
\mbox{Tr}\rho_A^n=\frac{\left|\lambda {\rm cn}(\lambda z_1|m){\rm dn}(\lambda z_1|m)\right|^{2\Delta_n}}{|2{\rm sn}(\lambda z_1|m)|^{2\Delta_n}}
\end{align}
and 
\begin{align}
S_n=-\frac{c}{12}\left(1+\frac{1}{n}\right)\log\xi.
\end{align}
where 
\begin{align}
\xi=\left[\frac{\lambda {\rm cn}(\lambda z_1|m){\rm dn}(\lambda z_1|m)}{|2{\rm sn}(\lambda z_1|m)|}\right],
\end{align}
with ${\rm sn}(z|m)$, ${\rm cn}(z|m)$ and ${\rm dn}(z|m)$ are Jacobi elliptic functions. 

We can use similar method to compute the mutual information dynamics. As shown in Fig.~\ref{fig:rectangle_MI}, we have
\begin{align}
&\frac{\mbox{Tr}\rho_{A\cup B}^n}{\mbox{Tr}\rho_{A}^n \mbox{Tr}\rho_{B}^n}=\frac{\langle \mathcal{T}_n(z_1,\overline{z}_1)\mathcal{T}_n(z_2,\overline{z}_2)\rangle_{\rm Rec}}{\langle \mathcal{T}_n(z_1,\overline{z}_1) \rangle_{\rm Rec} \langle \mathcal{T}_n(z_2,\overline{z}_2) \rangle_{\rm Rec}}\nonumber\\
 &=\frac{\langle \mathcal{T}_n(w_1,\overline{w}_1)\mathcal{T}_n(w_2,\overline{w}_2)\rangle_{\rm UHP}}{\langle \mathcal{T}_n(w_1,\overline{w}_1)\rangle_{\rm UHP} \langle \mathcal{T}_n(w_2,\overline{w}_2) \rangle_{\rm UHP}}\nonumber\\
 &=\frac{\langle \mathcal{T}_n(w_1)\mathcal{T}_n(\overline{w}_1)\mathcal{T}_n(w_2)\mathcal{T}_n(\overline{w}_2)\rangle_{\mathcal{C}}}{\langle \mathcal{T}_n(w_1)\mathcal{T}_n(\overline{w}_1)\rangle_{\mathcal{C}} \langle \mathcal{T}_n(w_2)\mathcal{T}_n(\overline{w}_2) \rangle_{\mathcal{C}}}\nonumber\\
 &=F(\eta)
\end{align}
where the  cross ratio $\eta$ is defined as
\begin{align}
\eta\equiv \frac{ |w_1-\overline{w}_1||w_2-\overline{w}_2|}{|w_1-\overline{w}_2||w_2-\overline{w}_1|}.
\end{align}
with $w_1={\rm sn}(\mathrm{i}\lambda L_A|m)$ and $w_2={\rm sn}(\mathrm{i}\lambda (L-L_B)|m)$.

\section{Master equation for continuous time model}
\label{app:mastereqn}
\subsection{Unitary Brownian dynamics}
We first consider the pure unitary evolution and take the Brownian evolution with 
\begin{align}
|\psi(t+dt)\rangle=e^{-idH}|\psi(t)\rangle
\end{align}
where $dH$ is a one dimensional random free fermion Hamiltonian, i.e.,
\begin{align}
dH=\sum_j (c_{j+1}^\dag c_j dW_j+ c_{j}^\dag c_{j+1} d\overline{W}_j).
\end{align}
$W_j(t)$ is Brownian motion with  
\begin{align}
dW_id\overline{W}_j=A\delta_{i,j}dt.
\end{align}
We compute the evolution of the $d|C_{a,b}|^2/dt$, which according to It$\hat{\rm o}$ calculus, should take the following form,
\begin{align}
d|C_{a,b}|^2= dC_{b,a} dC_{a,b}+\frac{d^2C_{b,a} C_{a,b}+C_{b,a} d^2C_{a,b}}{2}.
\end{align}
As we have shown in Eq.\eqref{eq:unitary_C}, the first derivative $dC$ satisfies,
\begin{align}
dC=i[dH, C].
\end{align}
Therefore we have
\begin{align}
&\frac{|dC_{a,b}|^2}{Adt}\nonumber\\
&=\left(|C_{b+1,a}|^2+|C_{b-1,a}|^2+|C_{b,a+1}|^2+|C_{b,a-1}|^2 \right)\nonumber\\
&-2\delta_{a,b+1}C_{b+1,b+1}C_{b,b}-2\delta_{a,b-1}C_{b-1,b-1}C_{b,b}.
\end{align}
The second derivative $d^2C$ is
\begin{align}
d^2 C=-[dH,[dH,C]],
\end{align}
which leads to
\begin{align}
&\frac{d^2C_{b,a} C_{a,b}+C_{b,a} d^2C_{a,b}}{2dt}\nonumber\\
&=-4A|C_{a,b}|^2+2A(C_{a+1,a+1}C_{a,a}+C_{a-1,a-1}C_{a,a})\delta_{a,b}.
\end{align}
Therefore we have
\begin{align}
&\frac{d|C_{a,b}|^2}{Adt}=\left(|C_{b+1,a}|^2+|C_{b-1,a}|^2+|C_{b,a+1}|^2+|C_{b,a-1}|^2 \right)\nonumber\\
&-2\delta_{a,b+1}C_{b+1,b+1}C_{b,b}-2\delta_{a,b-1}C_{b-1,b-1}C_{b,b}\nonumber\\
&-4|C_{a,b}|^2+2(C_{a+1,a+1}C_{a,a}+C_{a-1,a-1}C_{a,a})\delta_{a,b}.
\label{eq:unitary_eom}
\end{align}
It is easy to confirm that
\begin{align}
\sum_{a,b}\frac{d|C_{a,b}|^2}{dt}=0,
\end{align}
consistent with the constraint 
\begin{align}
\mbox{Tr}C^2=\sum_{a,b}|C_{a,b}|^2=N.
\end{align}
\subsection{Imaginary Brownian dynamics}
We consider the imaginary Brownian dynamics with 
\begin{align}
    |\psi(t+dt)\rangle \sim e^{-dH}|\psi(t)\rangle
\end{align}
where the Hamiltonian increment,
\begin{align}
dH=\sum_j c_{j}^\dag c_j dW_j\end{align}
with $dW$ satisfying
\begin{align}
dW_{i}dW_{j}=B\delta_{i,j}dt.
\end{align}
To compute the equation of motion for $C$ matrix, we need to compute $dC$ and $d^2C$. As we have shown in Eq.\eqref{eq:imaginary_C}, the first derivative satisfies 
\begin{align}
dC=-\{dH,C\}+2CdHC.
\end{align}
Therefore we have
\begin{align}
&\frac{|dC_{a,b}|^2}{Bdt}=2|C_{a,b}|^2+2\delta_{a,b}|C_{b,a}|^2\nonumber\\
&+4\sum_m|C_{b,m}|^2|C_{m,a}|^2-4|C_{a,b}|^2(C_{a,a}+C_{b,b}).
\end{align}
The second derivative has
\begin{align}
d^2C=&-dH\left[-\{dH,C\}+2CdHC\right]\nonumber\\
&-\left[-\{dH,C\}+2CdHC\right]dH\nonumber\\
&+2\left[-\{dH,C\}+2CdHC\right]dHC\nonumber\\
&+2CdH\left[-\{dH,C\}+2CdHC\right],
\end{align} 
with its matrix element
\begin{align}
&\frac{d^2C_{b,a}}{Bdt}=2C_{b,a}+2C_{b,a}\delta_{a,b}-4C_{b,a}(C_{b,b}+C_{a,a})\nonumber\\
&-4\sum_m C_{b,m}C_{m,a}+8\sum_m C_{b,m}C_{m,m}C_{m,a}.
\end{align}
Therefore we have
\begin{align}
&\frac{d^2C_{b,a} C_{a,b}+C_{b,a} d^2C_{a,b}}{2Bdt}\nonumber\\
&=2|C_{b,a}|^2+2|C_{b,a}|^2\delta_{a,b}-4|C_{b,a}|^2(C_{b,b}+C_{a,a})\nonumber\\
&-2\sum_m C_{b,m}C_{m,a}C_{a,b}-2\sum_m C_{a,m}C_{m,b}C_{b,a}\nonumber\\
&+4\sum_m C_{b,m}C_{m,a}C_{m,m}C_{a,b}+4\sum_m C_{a,m}C_{m,b}C_{m,m}C_{b,a}.
\end{align}
In total, we obtain
\begin{align}
&\frac{d|C_{a,b}|^2}{Bdt}=4|C_{a,b}|^2+4|C_{b,a}|^2\delta_{a,b}\nonumber\\
&-8|C_{b,a}|^2(C_{b,b}+C_{a,a})+4\sum_m|C_{b,m}|^2|C_{m,a}|^2\nonumber\\
&-2\sum_m\left[ C_{b,m}C_{m,a}C_{a,b}+ C_{a,m}C_{m,b}C_{b,a}\right]\nonumber\\
&+4\sum_m \left[C_{b,m}C_{m,a}C_{m,m}C_{a,b}+ C_{a,m}C_{m,b}C_{m,m}C_{b,a}\right].
\label{eq:imaginary_eom}
\end{align}

Notice that 
\begin{align}
&\sum_{a,b}\frac{dC_{b,a}dC_{a,b}}{Bdt}=2\sum_{a,b}|C_{a,b}|^2-2\sum_{a}|C_{a,a}|^2
\end{align}
and
\begin{align}
\sum_{a,b}\frac{dC_{b,a} C_{a,b}+C_{b,a}dC_{a,b}}{2Bdt}=-2\sum_{a,b}|C_{a,b}|^2+2\sum_{a}|C_{a,a}|^2.
\end{align}
Therefore we confirm 
\begin{align}
\sum_{a,b}\frac{d|C_{a,b}|^2}{dt}=0.
\end{align} 

\subsection{The unitary+imaginary Brownian dynamics}
We consider the mixed dynamics with
\begin{align}
    |\psi(t+dt)\rangle \sim e^{-idH}|\psi(t)\rangle
\end{align}
where the Hamiltonian increment is
\begin{align}
    dH=\sum_j\left(c_{j+1}^\dag c_j dW^1_j+c^\dag_j c_{j+1}d\overline{W}^1_j-ic^\dag_j c_j dW^2_j\right).
\end{align}
The equation of motion for $|C_{a,b}|^2$ is the combination of Eq.\eqref{eq:unitary_eom} and Eq.\eqref{eq:imaginary_eom}. We expect that by  solving this equation, we would obtain the same dynamics for the correlation function investigated in the main text. However, this equation is very complicated and it is hard to extract physics directly from it. 
We now derive an (approximate) master equation for $f_n$, as defined in (\ref{eq:fndef})

For the initial state with only diagonal element $C_{a,a}\neq 0$, we have $f_0=1$ and $f_{n>0}=0$. To study the dynamics of $f_n$, we start from Eq.\eqref{eq:unitary_eom} and Eq.\eqref{eq:imaginary_eom} and rewrite them in terms of $f_n$. For Eq.\eqref{eq:unitary_eom}, it mainly contributes a term 
\begin{align} f_{n-1}+f_{n+1}-2f_n,\end{align}
which is responsible for the diffusive spreading of $f_n$ if the dynamics is purely unitary. In Eq.\eqref{eq:imaginary_eom}, we throw away terms $C_{b,m}C_{m,a}C_{a,b}$ and $C_{b,m}C_{m,a}C_{m,m}C_{a,b}$ when $m\neq a$ or $b$. We make this approximation because under random dynamics, $\overline{C_{a,b}}=0$ if $a\neq b$ and their products are also zero. The term $\sum_m |C_{b,m}|^2|C_{m,a}|^2$ is very important and will contribute two quadratic terms:
\begin{align}
    \sum_{m=1}^\infty f_m f_{m+n} + \frac{1}{2}\sum_{m=1}^{n-1}f_m f_{n-m}.
\end{align}
The term $|C_{a,b}|^2C_{a,a}$ will contribute
\begin{align}
    f_n\sum_{m=1}^\infty f_m,
\end{align}
this is because $C_{a,a}=\sum_{b}|C_{a,b}|^2$. 
Including all these above terms, we can write down the non-linear master equation for $f_n$, which satisfies
\begin{subequations} \label{eq:master_app}\begin{align}
\frac{df_1}{dt} &=  \mu + \theta (f_2-2f_1) - 2f_1 \sum_{m=1}^\infty f_m  \notag \\
&+ \sum_{m=1}^\infty f_m f_{m+1}, \\
\frac{df_n}{dt} &= \theta(f_{n+1}+f_{n-1}-2f_n) - 2f_n \sum_{m=1}^\infty f_m \notag \\
&+ \sum_{m=1}^\infty f_m f_{m+n} + \frac{1}{2}\sum_{m=1}^{n-1}f_m f_{n-m}, \;\;\; (n>1)
\end{align}\end{subequations}
where $\mu$ and $\theta$ are positive constants. $\mu\sim \delta_{a,b+1}(C_{b,b}-C_{b+1,b+1})^2+\delta_{a,b-1}(C_{b,b}-C_{b-1,b-1})^2$ is the source term and is coming from the fluctuations of diagonal elements of $C$ matrix. In the above master equation, we do not consider the dynamics for $f_0$. This is because there is an extra constraint $\sum_{a}C_{a,a}=N$ and therefore $f_0$ cannot be simply described by the master equation. 

\bibliographystyle{apsrev4-1}
\bibliography{free_fermion}

\end{document}